\UseRawInputEncoding
\documentclass[twocolumn] 
{aastex631}

\usepackage{bm} 
\usepackage{enumitem}
\usepackage{amsmath}

\begin{document}

\title{ODIN: Improved Narrowband Ly$\bm{\alpha}$ Emitter Selection Techniques for $\bm{z}$ = 2.4, 3.1, and 4.5}



\author[0000-0002-9811-2443]{Nicole M. Firestone}
\affiliation{Department of Physics and Astronomy, Rutgers, the State University of New Jersey, Piscataway, NJ 08854, USA}

\author[0000-0003-1530-8713]{Eric Gawiser}
\affiliation{Department of Physics and Astronomy, Rutgers, the State University of New Jersey, Piscataway, NJ 08854, USA}

\author[0000-0002-9176-7252]{Vandana Ramakrishnan}
\affiliation{Department of Physics and Astronomy, Purdue University, 525 Northwestern Ave., West Lafayette, IN 47906, USA}

\author[0000-0003-3004-9596]{Kyoung-Soo Lee}
\affiliation{Department of Physics and Astronomy, Purdue University, 525 Northwestern Ave., West Lafayette, IN 47906, USA}

\author[0000-0001-5567-1301]{Francisco Valdes}
\affiliation{NSF’s National Optical-Infrared Astronomy Research Laboratory, 950 N. Cherry Ave., Tucson, AZ 85719, USA}

\author[0000-0001-9521-6397]{Changbom Park}
\affiliation{Korea Institute for Advanced Study, 85 Hoegi-ro, Dongdaemun-gu, Seoul 02455, Republic of Korea}

\author[0000-0003-3078-2763]{Yujin Yang}
\affiliation{Korea Astronomy and Space Science Institute, 776 Daedeokdae-ro, Yuseong-gu, Daejeon 34055, Republic of Korea}

\author[0000-0002-1328-0211]{Robin Ciardullo}
\affiliation{Department of Astronomy \& Astrophysics, The Pennsylvania
State University, University Park, PA 16802, USA}
\affiliation{Institute for Gravitation and the Cosmos, The Pennsylvania
State University, University Park, PA 16802, USA}


\author[0000-0003-0570-785X]{Mar\'ia Celeste Artale}
\affiliation{Universidad Andres Bello, Facultad de Ciencias Exactas, Departamento de Fisica, Instituto de Astrofisica, Fernandez Concha 700, Las Condes, Santiago (RM), Chile}

\author{Barbara Benda}
\affiliation{Department of Physics and Astronomy, Rutgers, the State University of New Jersey, Piscataway, NJ 08854, USA}
\affiliation{Department of Physics, University of Washington, Seattle, WA 98195, USA}

\author{Adam Broussard}
\affiliation{Department of Physics and Astronomy, Rutgers, the State University of New Jersey, Piscataway, NJ 08854, USA}

\author{Lana Eid}
\affiliation{Department of Physics and Astronomy, Rutgers, the State University of New Jersey, Piscataway, NJ 08854, USA}

\author{Rameen Farooq}
\affiliation{Department of Physics and Astronomy, Rutgers, the State University of New Jersey, Piscataway, NJ 08854, USA}

\author[0000-0001-6842-2371]{Caryl Gronwall}
\affiliation{Department of Astronomy \& Astrophysics, The Pennsylvania
State University, University Park, PA 16802, USA}
\affiliation{Institute for Gravitation and the Cosmos, The Pennsylvania
State University, University Park, PA 16802, USA}

\author[0000-0002-4902-0075]{Lucia Guaita}
\affiliation{Universidad Andres Bello, Facultad de Ciencias Exactas, Departamento de Fisica, Instituto de Astrofisica, Fernandez Concha 700, Las Condes, Santiago (RM), Chile}

\author[0000-0001-8221-8406]{Stephen Gwyn}
\affiliation{Herzberg Astronomy and Astrophysics Research Centre, National Research Council of Canada, Victoria, British Columbia, Canada}

\author[0000-0003-3428-7612]{Ho Seong Hwang}
\affiliation{Department of Physics and Astronomy, Seoul National University, 1 Gwanak-ro, Gwanak-gu, Seoul 08826, Republic of Korea}
\affiliation{SNU Astronomy Research Center, Seoul National University, 1 Gwanak-ro, Gwanak-gu, Seoul 08826, Republic of Korea}

\author[0009-0003-9748-4194]{Sang Hyeok Im}
\affiliation{Department of Physics and Astronomy, Seoul National University, 1 Gwanak-ro, Gwanak-gu, Seoul 08826, Republic of Korea}

\author[0000-0002-2770-808X]{Woong-Seob Jeong}
\affiliation{Korea Astronomy and Space Science Institute, 776 Daedeokdae-ro, Yuseong-gu, Daejeon 34055, Republic of Korea}

\author[0009-0002-6186-0293]{Shreya Karthikeyan}
\affiliation{Department of Astronomy, University of Maryland, College Park, MD 20742, USA}

\author[0000-0002-1172-0754]{Dustin Lang}
\affiliation{Perimeter Institute for Theoretical Physics, 31 Caroline Street North, Waterloo, ON N2L 2Y5, Canada}

\author[0009-0008-4022-3870]{Byeongha Moon}
\affiliation{Korea Astronomy and Space Science Institute, 776 Daedeokdae-ro, Yuseong-gu, Daejeon 34055, Republic of Korea}

\author[0000-0001-9850-9419]{Nelson Padilla}
\affiliation{Instituto de Astronom\'ıa Te\'orica y Experimental (IATE), CONICET-UNC, Laprida 854, X500BGR, C\'ordoba, Argentina}

\author[0000-0002-7712-7857]{Marcin Sawicki}
\affiliation{Institute for Computational Astrophysics and Department of Astronomy and Physics, Saint Mary’s University, 923 Robie Street, Halifax, Nova Scotia,
B3H 3C3, Canada}

\author[0009-0007-1810-5117]{Eunsuk Seo}
\affiliation{Department of Astronomy and Space Science, Chungnam National University, 99 Daehak-ro, Yuseong-gu, Daejeon, 34134, Republic of Korea}

\author{Akriti Singh}
\affiliation{Departamento de Ciencias Fisicas, Universidad Andres Bello, Fernandez Concha 700, Las Condes, Santiago, Chile}
\affiliation{European Southern Observatory Las Condes, Región Metropolitana, Chile}

\author[0000-0002-4362-4070]{Hyunmi Song}
\affiliation{Department of Astronomy and Space Science, Chungnam National University, 99 Daehak-ro, Yuseong-gu, Daejeon, 34134, Republic of Korea}

\author[0000-0001-6162-3023]{Paulina Troncoso Iribarren}
\affiliation{Escuela de Ingeniería, Universidad Central de Chile, Avenida Francisco de Aguirre 0405, 171-0164 La Serena, Coquimbo, Chile}


\begin{abstract}

Lyman-Alpha Emitting galaxies (LAEs) are typically young, low-mass, star-forming galaxies with little extinction from interstellar dust. Their low dust attenuation allows their Ly$\alpha$ emission to shine brightly in spectroscopic and photometric observations, providing an observational window into the high-redshift Universe. Narrowband surveys reveal large, uniform samples of LAEs at specific redshifts that probe large scale structure and the temporal evolution of galaxy properties. The One-hundred-deg$^2$ DECam Imaging in Narrowbands (ODIN) utilizes three custom-made narrowband filters on the Dark Energy Camera (DECam) to discover LAEs at three equally spaced periods in cosmological history. In this paper, we introduce the \textit{hybrid-weighted double-broadband continuum estimation} technique, which yields improved estimation of Ly$\alpha$ equivalent widths. Using this method, we discover 6032, 5691, and 4066 LAE candidates at $z =$ 2.4, 3.1, and 4.5 in the extended COSMOS field ($\sim$9 deg$^2$). We find that [\ion{O}{2}] emitters are a minimal contaminant in our LAE samples, but that interloping Green Pea-like [\ion{O}{3}] emitters are important for our redshift 4.5 sample. We introduce an innovative method for identifying [\ion{O}{2}] and [\ion{O}{3}] emitters via a combination of narrowband excess and galaxy colors, enabling their study as separate classes of objects. We present scaled median stacked SEDs for each galaxy sample, revealing the overall success of our selection methods. We also calculate rest-frame Ly$\alpha$ equivalent widths for our LAE samples and find that the EW distributions are best fit by exponential functions with scale lengths of $w_0$ = 53$\pm$1, 65$\pm$1, and 59$\pm$1 {\AA}, respectively. 





\end{abstract}


\NewPageAfterKeywords

\section{Introduction}\label{sec:intro}

The presence of significant Lyman Alpha (Ly$\alpha$) emission in young, star forming galaxies was first theorized by \citet{Partridge1967}. Today, we understand Ly$\alpha$ Emitting galaxies (LAEs) as young, low-mass, low-dust, star-forming systems, which have been identified as predecessors of Milky Way-type galaxies \citep[e.g.,][]{gawiser2007lyalpha, guaita2010lyalpha, walker2012present, pucha2022lyalpha}. LAEs have prominent Ly$\alpha$ emission due to the recombination of hydrogen in their interstellar media (ISM) and, in some cases, scattering that occurs in the circumgalactic medium (CGM). In the ISM, ionization is driven by active star formation \citep[specifically hot young O-type and B-type stars; e.g.,][]{kunth1998hst, hui1997equation} or the presence of an active galactic nucleus (AGN) \citep[e.g.,][]{padmanabhan2021distinguishing}. After ionization via either of the aforementioned processes, the Hydrogen undergoes recombination, producing Ly$\alpha$ radiation in significant quantities. Because LAEs are typically nearly dust-free \citep[e.g.,][]{weiss2021hetdex}, the Ly$\alpha$ emission line formed through these processes does not experience severe extinction from interstellar dust and stands out as a prominent spectral feature. In the range $2 \lesssim $ $z$ $\lesssim 5$, the expansion of the Universe redshifts this Ly$\alpha$ emission line feature from the rest-frame wavelength of 121.6 nm into the optical regime, making LAEs observable by ground-based telescopes. 

After many years of unavailing searches for the fabled LAEs of \citet{Partridge1967}, the development of more sensitive telescopes and wider-field detectors in the mid-1990s brought with it some of the first notable LAE surveys (see \citet{Ouchi_2020} for a comprehensive review). One of the earliest successful LAE surveys was the Hawaii Survey, which used the 10m Keck II Telescope to conduct narrowband and spectroscopic searches for high equivalent width LAEs at $3 < $ $z$ $ < 6$ \citep{Hu_1998}. A few years later, the Large-Area Lyman Alpha (LALA) survey used the CCD Mosaic camera at the 4 m Mayall telescope at Kitt Peak National Observatory and the low-resolution imaging spectrograph (LRIS) instrument at the Keck 10m telescope to discover and spectroscopically confirm $z = 4.5$ LAEs \citep{rhoads2000first}. Shortly thereafter, the Subaru Deep Survey was conducted using narrowband imaging at $z = 4.86$ on the 8.2m Subaru Telescope \citep{ouchi2003subaru}. Then, the Multiwavelength Survey by Yale-Chile (MUSYC) used the MOSAIC-II Camera at the CTIO 4m telescope \citep{gawiser2006multiwavelength} to study LAEs at $z = 2.1$ \citep{guaita2010lyalpha} and $z = 3.1$ \citep{gawiser2007lyalpha}. Subsequently, Lyman Alpha Galaxies in the Epoch of Reionization (LAGER) Survey used the Dark Energy Camera (DECam) at the CTIO 4m telescope to study cosmological reionization at $z$ $\sim$ $7$ \citep{zheng2017first, harish2022new}. In recent years, the Hobby-Eberly Telescope Dark Energy Experiment (HETDEX) has taken the lead on spectroscopic LAE surveys \citep{gebhardt2021hobby}. Currently, the largest published narrowband-selected LAE samples have been discovered by the Systematic Identification of LAEs for Visible Exploration and Reionization Research Using Subaru HSC (SILVERRUSH), which used data from the Hyper Suprime-Cam (HSC) Subaru Strategic Program to discover LAEs over a wide range of redshifts \citep{ouchi2018systematic, kikuta2023silverrush}. 

Large, uniform samples of LAEs have a wide range of uses for studies of galaxy formation, galaxy evolution, large scale structure, and cosmology. High-redshift LAEs ($z$ $\gtrsim$ 6) can be used to probe the epoch of cosmic reionization, the era in which the neutral matter that existed after the recombination became ionized by first generation stars \citep[e.g.,][]{steidel1999lyman, stark2010keck, schenker2014line, zheng2017first, Ouchi_2020, yoshioka2022chorus}. Additionally, LAEs serve as good tracers of the large scale structure of the Universe \citep[e.g.,][]{dey2016spectroscopic, shi2019galaxies, huang2022evaluating}, allowing us to study the temporal progression of the galaxy distribution at different epochs \citep[e.g.,][]{gawiser2007lyalpha, gebhardt2021hobby}. Since LAEs are composed of baryonic matter and dark matter halos, we can also use them as tools to measure the relationship between baryonic matter and dark matter, i.e., galaxy bias \citep{Coil_2013}. This type of analysis helps us to understand how high-redshift galaxies grow into the systems we see today \citep[e.g.,][]{gawiser2007lyalpha, ouchi2010statistics, guaita2010lyalpha}. Lastly, we can use LAEs to study star formation histories by fitting their rest-ultraviolet-through-near-infrared photometry \citep{iyer2019nonparametric, acquaviva2011spectral, acquaviva2011sed, iyer2017reconstruction}. This analysis allows us to characterize star formation episodes throughout the lifetime of galaxies, which can help us to better understand the physical processes that contribute to star formation and quenching in LAEs and how they compare to those in high-mass counterparts. Collectively, these scientific opportunities make LAEs a powerful observational tool for probing the high-redshift Universe, offering us many insights into the intricacies of galaxy formation and evolution, and cosmology. However, many of these studies require large, uniform samples of LAEs at well-separated periods in cosmological history. 

One-hundred-deg$^2$ DECam Imaging in Narrowbands (ODIN) is a 2021-2024 NOIRLab survey program designed to discover LAEs using narrowband imaging \citep{odin_survey, ramakrishnan2023odin}. ODIN's narrowband data is collected with DECam on the V\'ictor M. Blanco 4m telescope at the Cerro Tololo Inter-American Observatory (CTIO) in Chile. This project utilizes three custom-made narrowband filters with central wavelengths 419 nm (N419), 501 nm (N501), and 673 nm (N673) to create samples of LAE candidates during the period of Cosmic Noon at redshifts 2.4, 3.1, and 4.5, respectively. ODIN's narrowband-selected LAEs allow us to view large snapshots of the Universe 2.8, 2.1, and 1.4 billion years after the Big Bang, respectively. With ODIN, we expect to discover a sample of $>$100,000 LAEs in seven deep wide fields down to a magnitude of $\sim$25.7 AB, covering an area of $\sim$100 deg$^2$. ODIN’s carefully chosen filters and unprecedented number of LAEs will enable us to create and validate samples of the galaxy population at three equally spaced eras in cosmological history. Using these data, we can trace the large scale structure of the Universe, study the evolution of the galaxies' dark matter halo masses, and investigate the star formation histories of individual LAEs. 

In this paper, we introduce innovative techniques for selecting LAEs and reducing interloper contamination using ODIN data in the extended COSMOS field ($\sim$9 deg$^2$), and introduce ODIN's inaugural sample of $\sim$16,000 LAEs at $z$ = 2.4, 3.1, and 4.5. By generating this unprecedentedly large sample of LAEs with impressive sample purity, ODIN will be able to better understand galaxy formation, galaxy evolution, and the large scale structure of our Universe with significantly improved statistical robustness. From these results, we will be able to bind together chapters of the evolutionary biography of our Universe with what will be the largest sample of narrowband-selected LAEs to date. 

In Section \ref{sec:data} we discuss the data acquisition and preprocessing. In Section \ref{sec:el_gal_selection} we introduce the hybrid-weighted double-broadband continuum estimation technique and selection criteria for our emission line galaxy samples. In Section \ref{sec:results} we introduce our final emission line galaxy samples and discuss their scaled median stacked spectral energy distributions (SEDs) and emission line equivalent width distributions. In Section \ref{sec:conc} we outline our conclusions and future work. Throughout this paper, we assume $\Lambda$CDM cosmology with $h$ = 0.7, $\Omega_m$ = 0.27, and $\Omega_{\Lambda}$ = 0.73 and use comoving distance scales.

\section{Data}\label{sec:data}

\subsection{Images}

\begin{figure*}
\begin{center}
\includegraphics[width=0.9\textwidth]{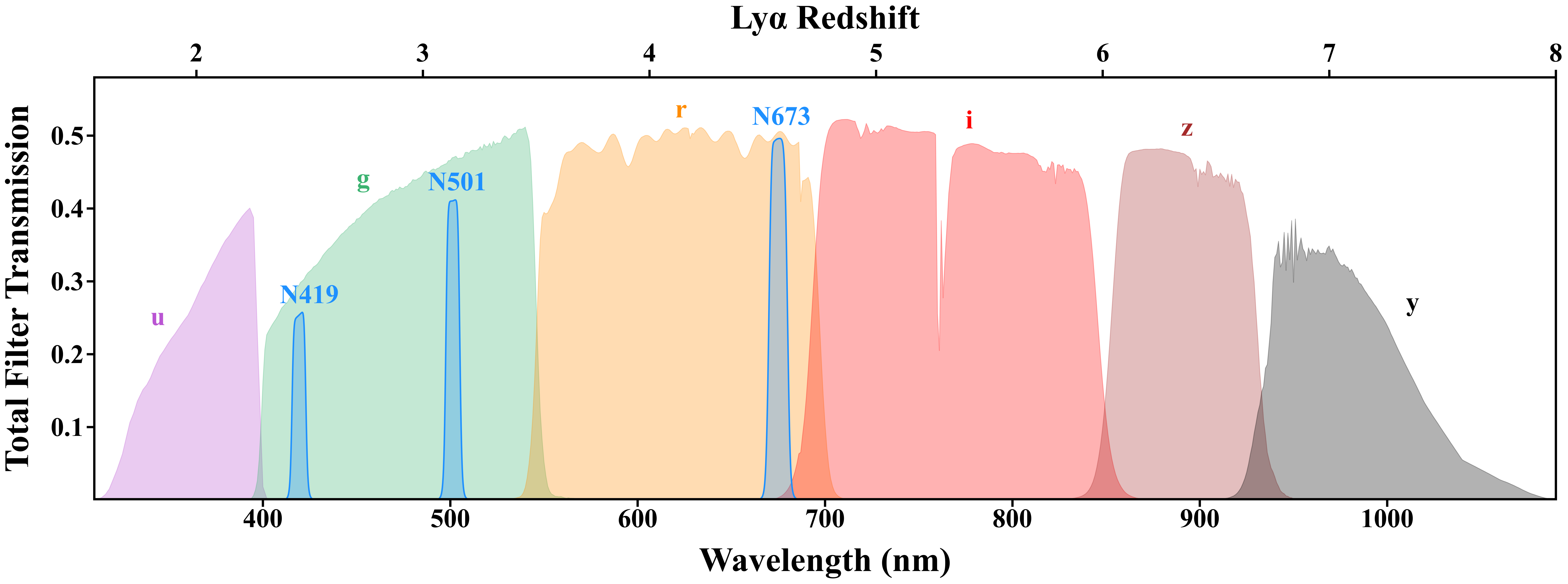}
\caption{Filter transmission for $N419$, $N501$, $N673$, $u$, $g$, $r$, $i$, $z$, and $y$-band filters as a function of wavelength (bottom axis) and Ly$\alpha$ redshift (top axis). The $u$, $g$, $r$, $i$, and $z$ band transmission curves are measured curves from CLAUDS and HSC \citep{CLAUDS, HSCSSP} while the $N419$, $N501$, and $N673$ transmission curves are simulated.} 
\label{fig:transmission_curves}
\end{center}
\end{figure*}

For ODIN's LAE selections, we require narrowband data as well as archival broadband data in the extended COSMOS field. The narrowband data for filters $N419$, $N501$, and $N673$ were collected using DECam on the Blanco 4m telescope at CTIO by the ODIN team \citep{odin_survey}. Archival $grizy$ broadband data were acquired from the Hyper Suprime-Cam Subaru Strategic Program (HSC-SSP) \citep{kawanomoto2018hyper, HSCSSP}. HSC-SSP data were collected using the wide-field imaging camera on the prime focus of the 8.2 m Subaru telescope \citep{HSCSSP}. HSC-SSP imaging in the COSMOS field includes two layers, Deep and Ultradeep \citep{HSCSSP}\null. Archival broadband data for the $u$-band were acquired from The CFHT Large Area $u$-band Deep Survey (CLAUDS) \citep{CLAUDS}. CLAUDS data were collected using the MegaCam mosaic imager on the Canada-France-Hawaii Telescope (CFHT) \citep{CLAUDS} and covered a smaller area than the HSC-SSP\null. The effective wavelength, seeing, depth, and extinction coefficients (see Section \ref{subsec:dust_coors}) in the COSMOS field for each filter are presented in Table \ref{tab:filters}. The $grizy$ \textit{seeing} is reported as the median seeing value for each COSMOS wide-depth stack. Since the COSMOS field includes two layers for the HSC broadband data, we present the parameters for both the Deep and Ultradeep regions separated by a slash when necessary. The transmission curves for all of these filters are presented in Figure \ref{fig:transmission_curves}. 




\begin{table}[h!]
\centering
\caption{Filter name, effective wavelength, full-width-half-max, seeing, depth, and extinction coefficient ($k$) for each filter in COSMOS \citep{odin_survey, CLAUDS, HSCSSP}.}\label{tab:filters}
\begin{tabular}{cccccc}
\hline\hline
Filter & $\lambda_{eff}$ & FWHM & Seeing & Depth & $k$ \\ 
 & (nm) & (nm) & (arcsec) & (mag) & \\ 
\hline
$N419$ & 419.3 & 7.5 & 1.1 & 25.5 & 3.64 \\ 
$N501$ & 501.4 & 7.6 & 0.9 & 25.7 & 3.03 \\ 
$N673$ & 675.0 & 10.0 & 1.0 & 25.9 & 2.01 \\ 
$u$ & 368.2 & 86.8 & 0.92 
& 27.7 
& 4.06 \\
$g$ & 481.1 & 139.5 & 0.74 & 27.8/28.4 & 3.17 \\
$r$ & 622.3 & 150.3 & 0.79 & 27.4/28.0 & 2.28 \\
$i$ & 767.5 & 157.4 & 0.57 & 27.1/27.7 & 1.61 \\
$z$ & 890.8 & 76.6 & 0.75 & 26.6/27.1 & 1.24 \\
$y$ & 978.5 & 78.3 & 0.73 & 25.6/26.6 & 1.09 \\
\hline
\end{tabular}
\end{table}



\subsection{Source Extractor Catalogs}

In order to carry out source detection, we first divide the narrowband stack into ``tracts'' to match the $grizy$ images from the HSC-SSP \citep{HSCSSP}. Each tract spans an area of $\sim$ 1.7 $\times$ 1.7 deg$^2$, with a small overlap between neighboring tracts. We select sources from each tract image separately using the Source Extractor (SE) software \citep{SExtractor} run in dual image mode with one narrowband image as the detection band and the $grizy$ plus remaining narrowband images as the measurement bands. This allows us to measure the source fluxes in identical apertures on all the frames. We measure the photometry in multiple closely spaced apertures, making it possible to interpolate the fractional flux enclosed within an aperture of any radius. While running SE, we filter each image with a Gaussian kernel with FWHM matched to the narrowband point spread function. We impose a detection threshold (\texttt{DETECT\_THRESH}) of 0.95$\sigma$, where $\sigma$ is the fluctuation in the sky value of the narrowband image, and a minimum area (\texttt{DETECT\_MINAREA}) of one pixel. These settings are optimized to detect faint point sources, which form the bulk of the LAE population. The specific value of \texttt{DETECT\_THRESH} is chosen to maximize the number of sources detected while still ensuring that the contamination of the source catalog by noise peaks remains below 1\%. The extent of the contamination is estimated by running SE on a sky-subtracted and inverted (``negative'') version of the narrowband image. In this negative image, any true sources will be well below the detection threshold; any objects detected by SE are thus the result of sky fluctuations. So long as the sky fluctuations are Gaussian, i.e. the extent of the fluctuations above the mean is the same as that below, the number of sources detected in the negative image will be comparable to the number of false source selected with a given detection threshold.

\begin{figure}[h]
\begin{center}
\includegraphics[width=0.48\textwidth]{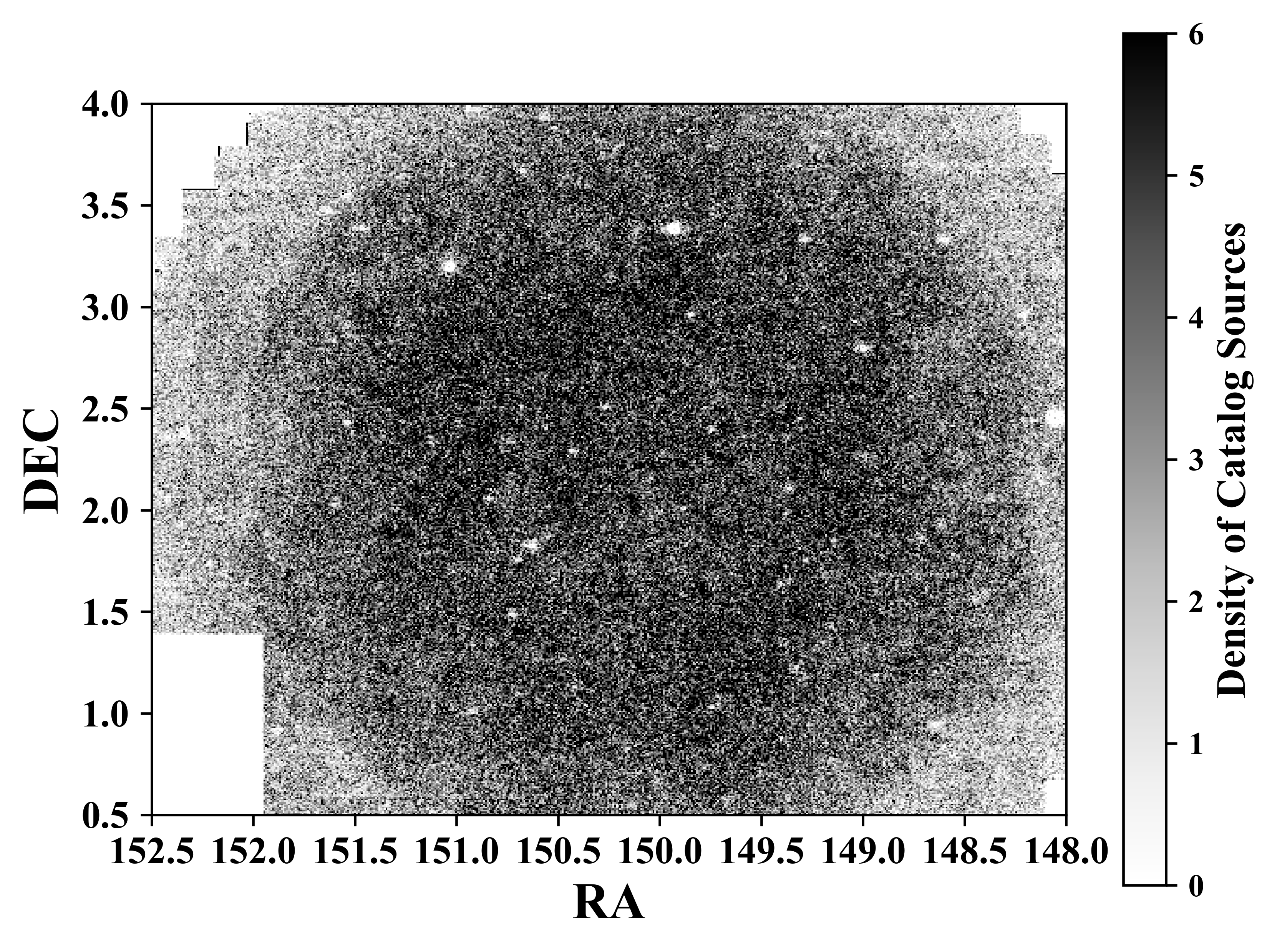}
\caption{2D histogram of COSMOS/$N419$ Source Extractor catalog. The x-axis represents the right ascension (RA) in degrees and the y-axis represents the declination (DEC) in degrees. The colorbar indicates the density of sources in each 2D bin. Only 9 deg$^2$ near the center, which has uniform depth, overlap with the HSC broad-band data.}
\label{fig:cat}
\end{center}
\end{figure}

The COSMOS/$N419$ SE catalog is presented in Figure \ref{fig:cat}. Note that this plot excludes regions where there is no overlap between the DECam and HSC-SSP/CLAUDS frames. After acquiring archival data and creating a source catalog, we carry out a series of steps related to data preprocessing, which are outlined in Subsections \ref{subsec:dust_coors} - \ref{subsubsec:starmasking}.





\subsection{Galactic Dust Corrections}\label{subsec:dust_coors}

As radiation from an extragalactic source travels through the Milky Way, it encounters dust clouds that cause absorption and scattering. As a consequence of this, the observed radiation from those sources appears to be dimmer and redder than the intrinsic radiation. In order to account for this effect and recover the intrinsic emission from the sources, we apply Galactic dust corrections to the data. 

We estimate the amount of reddening that a source experiences by comparing its observed B-V color to its intrinsic B-V color, i.e., E(B-V). In order to calculate the E(B-V) value for each of our sources, we use the reddening map of \citet[][hereafter SFD]{sfd_dust}, as modified by \citet{schlafly2011measuring}, along with an \citet{fitzpatrick1999correcting} reddening law. The resulting extinction coefficients for each filter, as interpolated from the DECam filter values presented by \citet{schlafly2011measuring}, are presented in Table \ref{tab:filters}. 

To implement Galactic dust corrections, we apply Equation \ref{eq:Gal_dust}, 
\begin{equation}\label{eq:Gal_dust}
    \text{flux}_{corr} = \text{flux}_{obs} \times 10^{0.4 k E(B-V)},
\end{equation} 
where $\text{flux}_{corr}$ represents the Galactic dust corrected flux value, $\text{flux}_{obs}$ represents the observed flux value, $k$ represents the extinction coefficient for a particular filter, and $E(B-V)$ represents the SFD reddening value for a particular source. 

At this point, we reassign low-flux density ($<3.6E-7$ $\mu Jy$) objects a magnitude of 40 as a flag, which is intentionally chosen to be much dimmer than the main distribution of magnitudes (centered around 24 for the full sample).

\subsection{Aperture Corrections For Photometry}

To fully account for the intrinsic brightness of each source, it is imperative that we also apply aperture corrections to our photometry. Each SE-generated source catalog produces flux density measurements for 12 different aperture diameters. Ideally, using the largest aperture available would yield the most accurate total flux measurements for the sources. However, by nature, the larger the aperture we use, the more noise is introduced by the background sky. On the other hand, if we use a smaller aperture we will underestimate the total flux densities of the sources but reduce the noise in our data. In order to accurately report the flux densities of our sources \textit{and} limit the noise in the data, we use smaller apertures for the flux density measurements of the sources and apply correction factors to estimate the total flux density of a source in each filter. These flux density corrections are also carried through to the magnitude values and all errors. In order to properly treat point sources and extended sources, we use slightly different methodology for each class of objects. 

\begin{figure}[h]
\begin{center}
\includegraphics[width=0.40\textwidth]{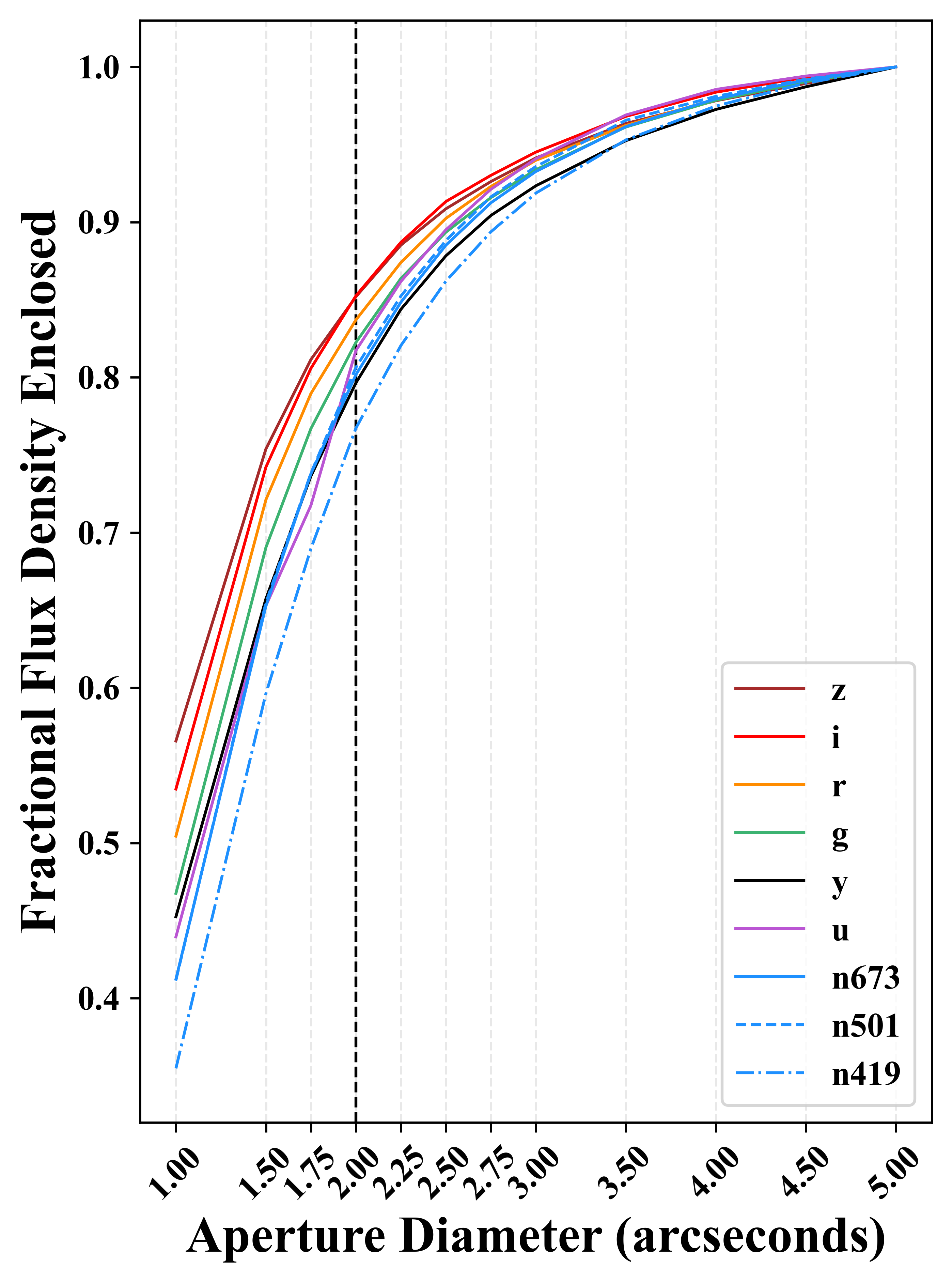}
\caption{Curves of Growth for the $u$, $g$, $r$, $i$, $z$, $y$, $N419$, $N501$, and $N673$ filters. The x-axis represents the aperture diameters in arcseconds and the y-axis represents the fractional flux density enclosed with respect to the largest aperture (5 arcseconds). The darker vertical line corresponds to the chosen aperture diameter of 2 arcseconds. The order of elements in the legend corresponds to the respective (vertical) fractional flux density in the 1 arcsecond aperture.}
\label{fig:cog}
\end{center}
\end{figure}

To produce aperture correction factors for point source, we examine the 2D integral of the point spread function (which we will henceforth refer to as the Curve of Growth) for each filter (see Figure \ref{fig:cog}). The Curves of Growth are constructed by plotting the median fractional flux density enclosed with respect to the largest aperture (5.0 arcsecond diameter) $frac\text{ }flux_n = \frac{f_n}{f_5}$ for bright, unsaturated point sources as a function of aperture diameter for each filter. We classify bright, unsaturated point sources as sources that obey the following criteria: 
\begin{itemize} [leftmargin=1.5\parindent]
  \item Respective magnitude between 18 and 19 (\textbf{bright})
  \item \texttt{FLAGS} $<$ 4 (\textbf{unsaturated}) 
  \item \texttt{FLUX\_RADIUS} $\leq$ 0.85 arcsecond (\textbf{point source})
\end{itemize}
We choose the bright source magnitude range by finding the magnitudes for which the median fractional flux density levels out and the normalized median absolute deviation (NMAD) of the fractional flux density is close to zero in all filters. We choose to use the NMAD rather than the standard deviation because the NMAD is less sensitive to outliers. In order to omit sources with pixel saturation, we include only sources with the SE \texttt{FLAGS} parameter $<$ 4. For the purpose of aperture corrections, we treat objects with a half-light radius (\texttt{FLUX\_RADIUS}) less than or equal to 0.85 arcseconds as point sources. 

After creating Curves of Growth with the subset of sources that obey these criteria, we convert the median fractional flux density for a particular aperture into a correction factor for each filter $corr = {1}/({frac\text{ }flux_n})$, such that $corr \times f_{n} = f_{5}$. The Curves of Growth for the COSMOS/$N419$ SE catalog are presented as a representative example in Figure \ref{fig:cog}. As a supplemental test of robustness, we also ensure that the Curve of Growth for each filter does not change dramatically across the survey area. 



To produce aperture correction factors for extended sources, we perform a regression analysis to determine a correction factor as a function of source half-light radius in the chosen 2 arcsecond aperture for each filter. This step allows us to limit contamination from uncorrected extended sources in the candidate sample. 


Ultimately, implementing these aperture corrections allows us to use a smaller aperture to better estimate the total flux of point sources without significantly biasing extended sources, while keeping the noise lower in the data. Additionally, at this step we apply a magnitude (flux) reassignment of 40 to sources whose flux values are low (including negative).  

\subsection{Starmasking}\label{subsubsec:starmasking}

\begin{figure*}
\begin{center}
\includegraphics[width=0.9\textwidth]{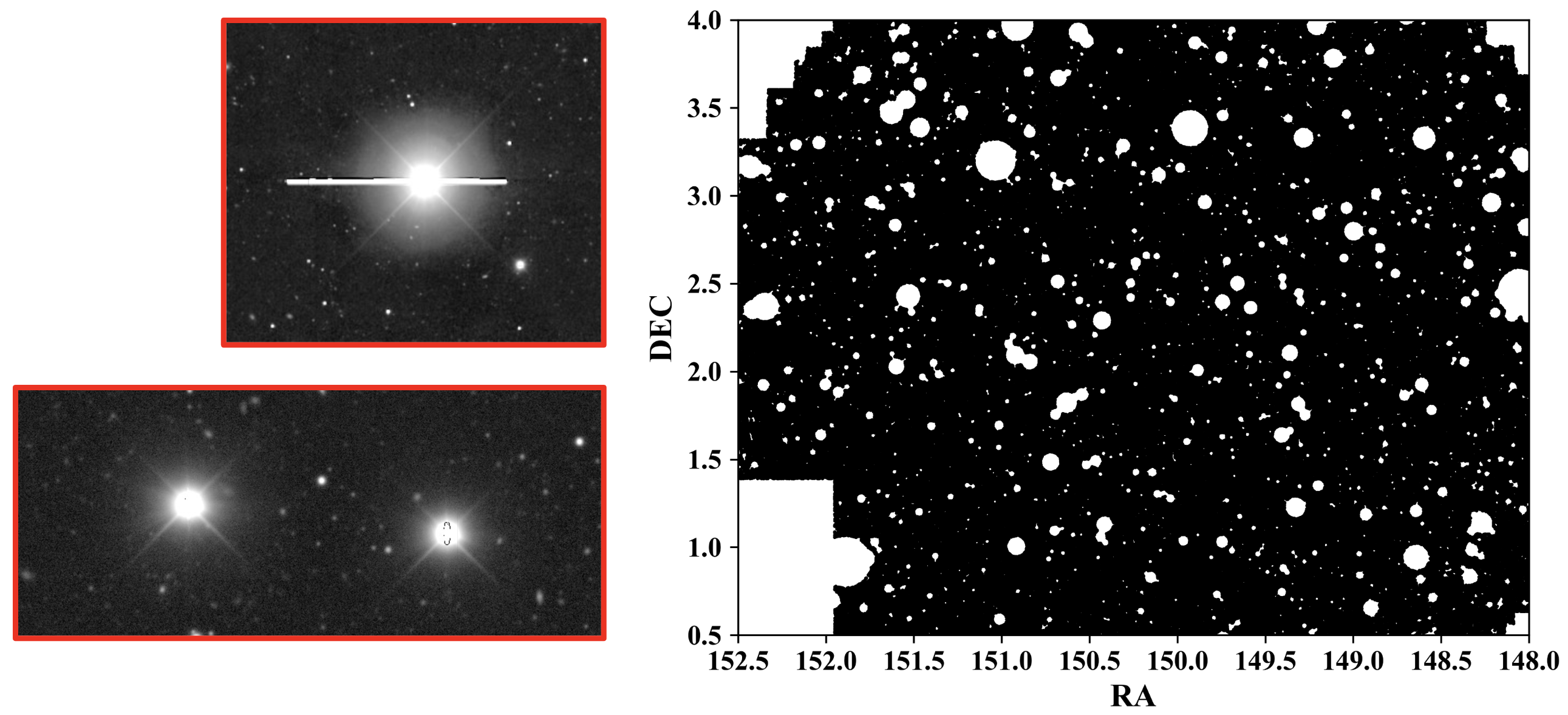}
\caption{Representative example of starmasking to eliminate CCD blooming and saturated stars from COSMOS/$N419$. In the upper left and lower left panels, we present examples of CCD blooming and saturated stars in ODIN's COSMOS/$N419$ images, respectively. In the right panel, the x-axis represents the right ascension (RA) in degrees and the y-axis represents the declination (DEC) in degrees. The sources that survive starmasking are presented in black.} 
\label{fig:starmasked_cat}
\end{center}
\end{figure*}

The next step in the candidate selection pipeline is starmasking. Starmasking removes data that have been contaminated by saturated stars and the effects of pixel oversaturation in the camera (CCD blooming). Starmasks were obtained from HSC-SSP \citep{starmasks}. We choose to use the $g$-band starmasks for this analysis because the individual masks were sufficiently sized for the narrowband images and did not have spurious objects. Examples of CCD blooming and saturated stars from the COSMOS/$N419$ sample as well as a visualization of the SE catalog after starmasking are presented in Figure \ref{fig:starmasked_cat} for reference. 

\subsection{Data Quality Cuts}

At this point, we apply data quality cuts in order to eliminate any poor or problematic data that are not accounted for in the starmasks. 

\begin{itemize} [leftmargin=1.5\parindent]

\item {$\bm{f_{\nu} \neq 0}$}
\\
We ensure that flux density $f_{\nu}$ for each source is nonzero in the narrowband and broadband filters chosen for each LAE selection (see Subsections \ref{subsec:N419}-\ref{subsec:N673} for details). This allows us to exclude sources with incomplete data.

\item $\bm{S/N_{NB} \geq 5}$
\\
We require that the narrowband signal to noise ratio for a source is greater than or equal to 5, where the signal to noise ratio is taken as the ratio of the narrowband flux density and the narrowband flux density error. This eliminates sources that should not have entered the SE catalogs.

\item $\bm{\texttt{IMAFLAGS\_ISO} = 0}$
\\
We require that the SE parameter \texttt{IMAFLAGS\_ISO} is equal to 0. \texttt{IMAFLAGS\_ISO} is a binary parameter, so a value of 0 indicates that all the pixels within a source's aperture have valid values and are unflagged, as opposed to a value of 1 indicating that any one pixel has no data or bad data in the external flag map \citep{SExtractor}.

\item $\bm{\texttt{FLAGS} < 4}$
\\
Lastly, we require that the SE \texttt{FLAGS} parameter is less than 4. This allows us to include sources whose aperture photometry is contaminated by neighboring sources and/or sources that had been deblended, and omit sources with pixel saturation \citep{SExtractor}.
 
\end{itemize}

\section{Emission Line Galaxy Selection}\label{sec:el_gal_selection}

\subsection{Improved Continuum Estimation Technique}



By definition, a true LAE has excess Ly$\alpha$ emission when compared with expected continuum emission at the Ly$\alpha$ wavelength. In order to select LAE candidates, we utilize narrowband and broadband filters to infer the presence of an emission line at the redshifted Ly$\alpha$ wavelength by looking for excess flux density in the narrowband. In order to measure this excess, we use a narrowband filter to capture the Ly$\alpha$ emission line and \textit{two} broadband filters to estimate the continuum emission at the narrowband effective wavelength. If a source's narrowband magnitude at this wavelength is significantly greater than the double-broadband continuum estimate, then the source is an LAE candidate.  


We estimate the continuum at the narrowband wavelength using two broadband filters by generating a weight for each filter according to Equation \ref{eq:weight_double_BB_1}, 
\begin{equation}\label{eq:weight_double_BB_1}
    \lambda_{NB} = w \lambda_a + (1-w) \lambda_b,
\end{equation}
where $\lambda$ represents the effective wavelength of a filter, $w$ is a weight, $NB$ represents the narrowband filter, and `a' and `b' generically represent two broadband filters. 
Since the effective wavelengths of each broadband filter are used to solve for $w$, $w$ will take on a value between 0 and 1 when used for an interpolation but can be outside that range when extrapolation is needed. 

In order to use these weights to generate a double-broadband continuum estimation, we begin by making the realistic assumption that continuum-only sources' have a power law flux distribution. In practice, this allows us to compute the double-broadband magnitude by linearly weighting the magnitude from each broadband filter. This weighted magnitude model is presented in Equation \ref{eq:power_law_flux}, where mag$_a$ is the magnitude in the `a' broadband filter, mag$_b$ is the magnitude in the `b' broadband filter, and $ab$ is the `ab' double-broadband continuum magnitude at the effective wavelength of the narrowband.
\begin{equation}\label{eq:power_law_flux}
    ab = (w) \text{mag}_a + (1-w) \text{mag}_b
\end{equation} However, when the noise is a key contributor, magnitudes become too unstable to get a reliable fit. We remedy this issue by using a simpler model for the subset of sources with low S/N in the flux density ($<10$\% of the starmasked source catalog). For this model, we assume that continuum-only sources’ flux density has a linear relationship to wavelength (as used in \citet{gawiser2006physical}). This weighted flux density model is presented in Equation \ref{eq:linear_flux}, where $f_a$ is the flux density in the `a' broadband filter, $f_b$ is the flux density in the `b' broadband filter, and $f_{ab}$ is the `ab' double-broadband continuum flux density at the effective wavelength of the narrowband.
\begin{equation}\label{eq:linear_flux}
    f_{ab} = (w) f_a + (1-w) f_b
\end{equation}

We refer to this new method as \textit{hybrid-weighted double-broadband continuum estimation}, in which we 
\begin{itemize}[leftmargin=1.5\parindent]
  \item Treat sources with S/N $\geq$ 3 in both single broadbands by assuming a power law flux density (i.e., weighted magnitude model; Equation \ref{eq:power_law_flux})
  \item Treat sources with S/N $<$ 3 in either broadband by assuming a linear flux density (i.e., weighted flux model; Equation \ref{eq:linear_flux})
\end{itemize}
For each sample, we use broadbands and weights $w$ according to Table \ref{tab:weights}.
\begin{table}[h!]
\centering
\caption{Double-broadband filter choices ($a$ and $b$) and corresponding weights ($w$ and $(1-w)$, respectively) for continuum estimation at each narrowband wavelength.}\label{tab:weights}
\begin{tabular}{ccccc}
\hline\hline
Narrowband & $a$ & $w$ & $b$ & $(1-w)$ \\ 
\hline
$N419$ & $r$ & $-0.438$ & $g$ & $1.438$\\
$N501$ & $g$ & $0.856$ & $r$ & $0.144$\\
$N673$ & $g$ & $0.323$ & $i$ & $0.677$\\
\hline
\end{tabular}
\end{table}
After applying this method, we implement a global narrowband zero point correction by adjusting the narrowband photometry such that the median narrowband excess is equal to zero for continuum-only objects. This correction is small and generally less than 10\%. 

This new method has many advantages for ODIN's datasets. First, it allows a better estimate the narrowband excess (equivalent width) of sources than is possible with a single broadband or flux density weighted double-broadband method. This is particularly advantageous for capturing dim LAEs. Additionally, it allows us to more effectively eliminate low redshift interlopers from the high redshift LAE candidates with minimal additional color cuts (see Subsections \ref{subsec:N673} and \ref{subsec:N501}). And lastly, it allows us to successfully use extrapolation (rather than interpolation) to estimate the continuum, which was not successful with a flux density weighted double-broadband method. This makes it possible to avoid direct use of the $u$-band filter for the $z$ = 2.4 LAE selection, which covers a smaller area and has more complex systematics than the $g$ and $r$ broadband filters (see Subsection \ref{subsec:N419}). Therefore, the improved hybrid-weighted double-broadband continuum estimation technique allows us to reduce interloper contamination and select candidates over a larger area with more robust photometry.

\subsection{LAE Selection Criteria}\label{subsec:sel_criteria}



Using hybrid-weighted double-broadband continuum estimation, we apply the following selection criteria to isolate LAEs:

\begin{enumerate} [leftmargin=1.3\parindent] 
    \item $\bm{(ab - NB) \geq (ab - NB)_{min}}$ 
\\
We require the narrowband excess of the LAE candidates to exceed an equivalent width cut according to Equation \ref{eq:ew_cut}, where $\lambda_{NB}$ is the effective wavelength of the narrowband filter, $\lambda_{Ly\alpha}$ is the minimum rest-frame wavelength of the Ly$\alpha$ emission line, $FWHM_{NB}$ is the full width at half maximum (FWHM) of the narrowband filter, and $EW_0$ is the rest-frame equivalent width of the Ly$\alpha$ emission line (which we take to be 20 $\text{\AA}$).
\begin{equation}\label{eq:ew_cut} 
(ab - NB)_{min} = 2.5\log_{10}\left[1+EW_0\left(\frac{\lambda_{NB}/\lambda_{Ly\alpha}}{FWHM_{NB}}\right)\right]
\end{equation}
For the $N419$, $N501$, and $N673$ narrowband filters, this equivalent width cut corresponds to narrowband excesses of 0.71, 0.83, and 0.82 magnitudes, respectively. In Section \ref{subsec:ew_estimations}, we will discuss a more complex process for equivalent width estimation based on these values. This cut allows us to limit the number of low-redshift interlopers that have other emission lines in the narrowband filters. This cut is quite robust to small-equivalent width interlopers, such as [\ion{O}{2}] emitting galaxies \citep{ciardullo2013hetdex}, though some Green Pea-like [\ion{O}{3}] emitters and AGNs may still remain in the sample (see Subsections \ref{subsec:N673}-\ref{subsec:N419}).
    \item $\bm{(ab - NB) \geq 3\sigma_{(ab - NB)}}$
\\
We require that candidates have a robust narrowband excess in order to avoid continuum-only objects being included due to the photometric uncertainties. Here, $\sigma_{(ab - NB)}$ is calculated by propagating the errors in $ab$ and $NB$.  
    \item $\bm{(BB - NB)} < \bm{-2.5\log_{10}{ \left[ \frac{C_{NB}}{C_{BB}} \right] + 2\sigma_{(BB-NB)} } }$
\\ 
We require that an object is at least as bright in the emission-line contributed broadband ($BB$) as a pure-emission-line LAE (infinite EW) would be, within 2$\sigma$ given possible noise fluctuations. Here, $C$ is given by Equation \ref{eq:C},
\begin{equation}\label{eq:C}
    C = \frac{\int{(c/\lambda^2)Td\lambda}}{T_{EL}},
\end{equation}
where $T$ is the filter transmission as a function of wavelength and $T_{EL}$ is obtained by averaging the filter transmission over the narrowband filter transmission curve, which is used as a proxy for the LAE redshift probability distribution function.
   \item $\bm{R_{50} < 1.38 ''}$
\\
We apply a cut in half-light radius $R_{50}$ to exclude large, extended sources. We define this limit as twice the NMAD in the half-light radii for sources that satisfy the above criteria from the half-light radii of bright, unsaturated point sources. This allows us to eliminate highly extended low-redshift contaminants whose photometry is not sufficiently corrected to avoid spurious narrowband excess. 
    \item $\bm{NB \geq 20}$
\\
We exclude sources with narrowband magnitude brighter than 20 in order to eliminate extremely bright contaminants, typically quasars or saturated stars. 
    \item $\bm{NB < D_{NB, 5\sigma} }$
\\
We eliminate spurious objects whose narrowband magnitude is dimmer than the median $5\sigma$ depth of the narrowband image $D_{NB, 5\sigma}$. For the $N419$, $N501$, and $N673$ narrowband filters, this magnitude corresponds to 25.5, 25.7, and 25.9 AB, respectively.
     \item $\bm{ \left|f_{H1}-f_{H2}\right| < 3 \sigma_{\left|f_{H1}-f_{H2}\right|} }$
\\
We divide the individual narrow-band images into two sets and use each to create a ``half-stack.''  We then eliminate objects whose flux density in these half-stacks ($f_{H1}$, $f_{H2}$) shows a statistically significant difference, since such objects are likely spurious. We determine the uncertainty of this difference $\sigma_{\left|f_{H1}-f_{H2}\right|}$ by summing the half-stack flux density errors in quadrature, $\sqrt{ \sigma_{H1}^2 + \sigma_{H2}^2}$.
 
\end{enumerate}

We also remove objects that appear in multiple LAE samples ($\sim$1\%), as these are likely to be bright low-$z$ interlopers such as AGNs. Finally, we apply additional color cuts to some of our LAE samples, which are designed to eliminate the largest known remaining sources of contamination in each dataset and enhance the purity of our LAE samples. The sources of contamination and cuts as well as the double-broadband choices for each filter set are described below.  

\subsection{Selection of $z = 4.5$ LAEs, $z = 0.81$ [\ion{O}{2}] Emitters, and  $z = 0.35$ [\ion{O}{3}] Emitters}\label{subsec:N673}

Out of the three samples of LAE candidates, the $N673$ catalog is the most susceptible to low redshift emission line galaxy interlopers. This is because the EW distributions and luminosity functions of low redshift interlopers climb as a function of redshift. The two most notable interlopers are $z$ $\approx$ 0.81 [\ion{O}{2}] emitters and $z$ $\approx$ 0.35 [\ion{O}{3}] emitters, with the most challenging culprit being the [\ion{O}{3}] emitters since the [\ion{O}{3}] emission line(s) tend to have larger EWs than the [\ion{O}{2}] emission line. We choose our selection filters specifically to isolate and remove these interlopers with minimal color cuts. 

\begin{figure*}
\begin{center}
\includegraphics[width=1\textwidth]{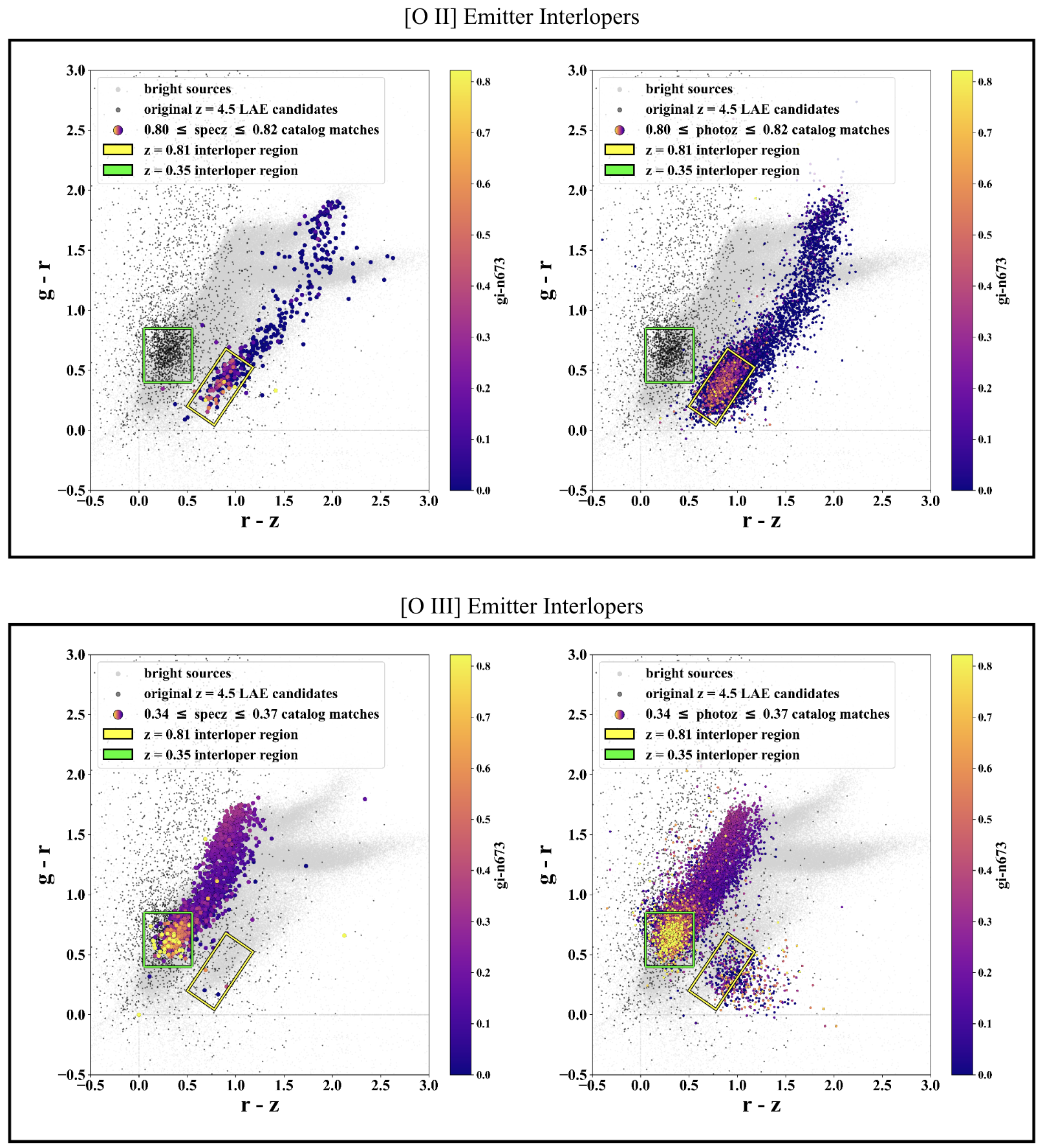}
\caption{$grz$ color-color diagrams demonstrating the compact, distinct regions in which [\ion{O}{2}] and [\ion{O}{3}] emitter interlopers reside. The top panel represents matches in the spectroscopic (left) and photometric (right) redshift source catalogs in the redshift range consistent with [\ion{O}{2}] emitter interlopers. The bottom panel contains the equivalent diagrams for [\ion{O}{3}] galaxies. The x-axis represents the difference between the $r$-band magnitude and the $z$-band magnitude, and the y-axis represents the difference between the $g$-band magnitude and the $r$-band magnitude. The light gray points represent bright objects in the SE source catalog. (Note that sources with a magnitude reassignment of 40 in at least two of the $g$, $r$, $z$-bands make up the light gray horizontal and vertical lines that intersect at (0, 0). These particular objects are bright in the $N673$ narrowband, but have low flux in the $g$, $r$, and $z$-bands.) The dark gray objects represent the original $z = 4.5$ LAE candidates prior to interloper rejection. The color bar shows the coding for the $(gi - N673)$ estimated EW of each cross-matched source. The yellow box shows the [\ion{O}{2}] emitter selection region, and the green box gives the locus of the [\ion{O}{3}] emitters.} 
\label{fig:specz_photoz_interlopers}
\end{center}
\end{figure*}

\begin{figure*}
\begin{center}
\includegraphics[width=1\textwidth]{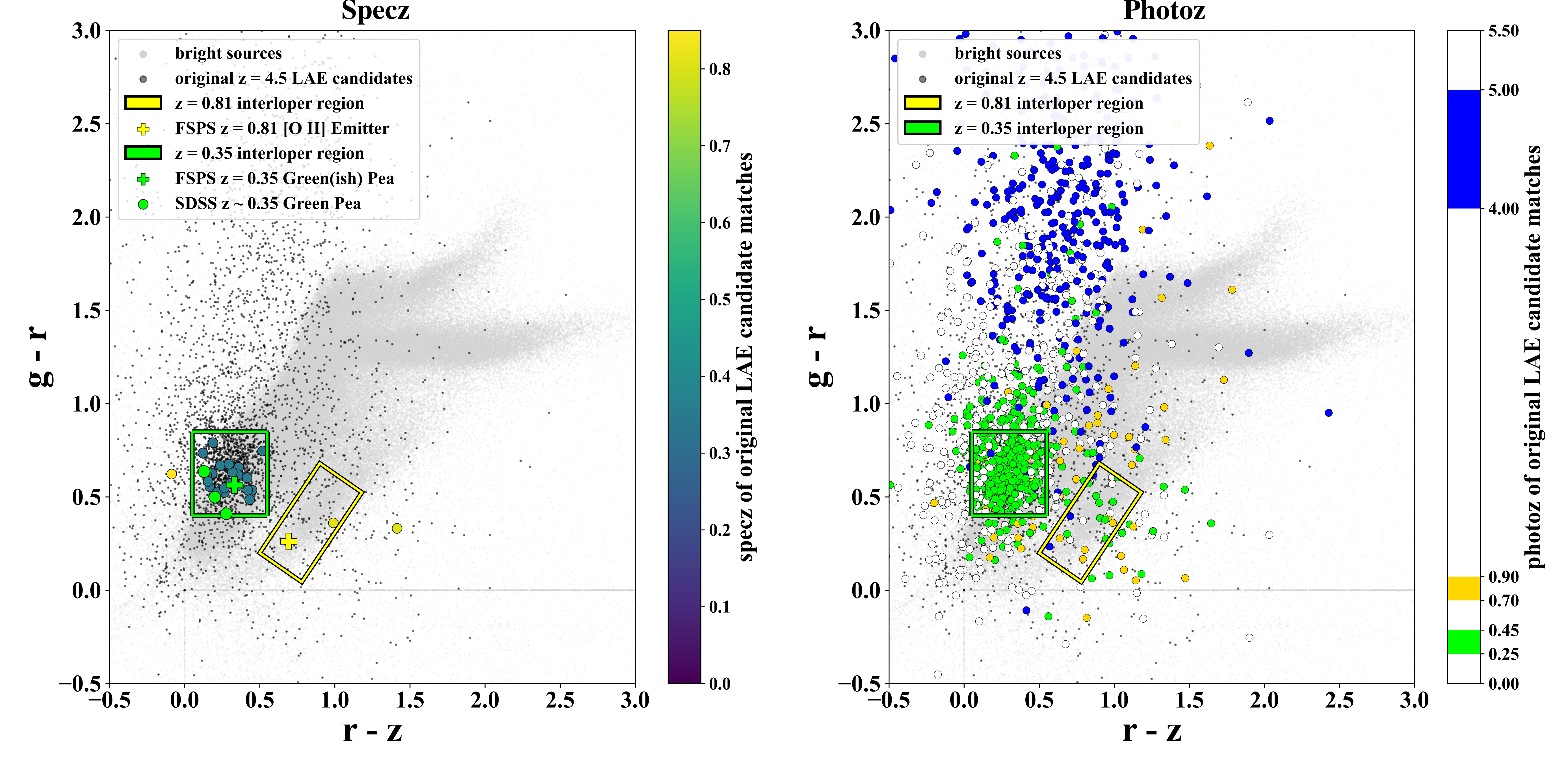}
\caption{$grz$ color-color diagrams demonstrating the distributions of spectroscopic and photometric redshift matches in the original $z = 4.5$ LAE candidate list prior to interloper rejection. Also shown are the positions of simulated interlopers and SDSS Green Peas. The general format of these plots follows that of Figure \ref{fig:specz_photoz_interlopers}. Here, the left panel introduces spectroscopic redshift matches in the original $z = 4.5$ LAE candidate sample and the color bars give the redshift coding of the matched objects. Additionally, the yellow and green plus signs in the left panel show the positions of an FSPS simulated $z = 0.81$ [\ion{O}{2}] emitter and a $z = 0.35$ Green Pea-like [\ion{O}{3}] emitter, respectively \citep{FSPS1, FSPS2}. The green circles represent $z$ $\sim $ 0.35 SDSS Green Peas \citep{galaxy_zoo}. The right panel introduces photometric redshift matches in the original $z = 4.5$ LAE candidate sample. The color bar distinguishes the redshift ranges corresponding to $z = 0.35$ [\ion{O}{3}] emitters (green), $z = 0.81$ [\ion{O}{2}] emitters (yellow), and $z = 4.5$ LAEs (blue). These diagrams reveal the presence of significant contamination from $z = 0.35$ [\ion{O}{3}] emitters in the original $z = 4.5$ LAE candidate sample. 
} 
\label{fig:specz_photoz_interlopers_cands}
\end{center}
\end{figure*}

For our $z = 4.5$ LAE selection, we carry out hybrid-weighted double-broadband continuum estimation using the $N673$, $g$-band, and $i$-band filters (see Figure \ref{fig:color_mag_comp} and Table \ref{tab:selection_results}). Following Table \ref{tab:weights}, we define the double-broadband continuum estimation $gi = 0.323 g + 0.677 i$. This combination of filters has significant advantages over using just $N673$ and the $r$-band. With the latter filter combination, not only do we have excess amounts of contamination from [\ion{O}{2}] and [\ion{O}{3}] emitters, but we do not capture all dim LAE candidates. Since N673 is the only one of our three narrowband filters for which it is feasible to perform interpolation between two broadband filters without either of them being affected by the Ly$\alpha$ emission line, we also explore this method. We find that excluding the $r$-band filter containing the emission line and instead using both $g$-band and $i$-band increases the number of dim LAE candidates selected. We also find that this choice of filter reduces contamination from lower EW [\ion{O}{2}] emitters and that the majority of our resulting contamination is from Green Pea-like [\ion{O}{3}] emitters (see Figures \ref{fig:specz_photoz_interlopers} and \ref{fig:specz_photoz_interlopers_cands}). Green Pea galaxies are compact extremely star-forming galaxies that are often thought of as low-$z$ LAE analogs \citep{galaxy_zoo}. 


In order to identify likely $z = 0.81$ [\ion{O}{2}] emitter and $z = 0.35$ [\ion{O}{3}] emitter interlopers in our data, we first carry out cross matches between the SE source catalog and archival spectroscopic/photometric redshift catalogs as well as between the initial LAE candidate catalogs and archival spectroscopic/photometric redshift catalogs. We obtain archival spectroscopic redshift data from \citet{skelton20143d, brammer20123d, silverman2015fmos, kashino2019fmos, coil2011prism, cool2013prism, bradshaw2013high, mclure2013sizes, maltby2016identification, scodeggio2018vimos, lefevre2003commissioning, le2005vimos, garilli2008vimos, cassata2011vimos, fevre2013vimos, drinkwater2018wigglez, lilly2007zcosmos, lilly2009zcosmos, kollmeier2017sdss} and we obtain photometric redshifts from \citet{weaver2022cosmos2020}.












As illustrated in Figure \ref{fig:specz_photoz_interlopers}, we find that objects in our source catalog that are matched to low-redshift $z = 0.81$ [\ion{O}{2}] emitters and $z = 0.35$ [\ion{O}{3}] emitters reside in specific, disjoint regions of $grz$ color-color space. Furthermore, we find that the sources in these redshift ranges with higher estimated ($gi - N673$) equivalent widths occupy compact and distinct regions of $grz$ color-color space. This can be seen in Figure \ref{fig:specz_photoz_interlopers}, where the colorbar displays the estimated narrowband excess from 0 to the $z = 4.5$ LAE EW cutoff. In addition to examining the ($gi - N673$) excess of the objects, we also examine their ($gr - N501$) values (see Figure \ref{fig:oiii_diagnostic}). Examining both of these excesses is helpful because ODIN's survey design ensures that the majority of $z = 0.35$ galaxies emitting Oxygen will have an [\ion{O}{3}] emission line in the $N673$ filter \textit{and} an [\ion{O}{2}] emission line in the $N501$ filter. We find that the objects with the highest ($gr - N501$) color are also concentrated in the region where we predicted significant contamination from $z = 0.35$ galaxies (see Figure \ref{fig:oiii_diagnostic}). This allows us to see that LAE selections at this redshift are strongly susceptible to $z = 0.35$ [\ion{O}{3}] emitter interlopers and mildly susceptible to $z = 0.81$ [\ion{O}{2}] emitter interlopers. 

\begin{figure*}
\begin{center}
\includegraphics[width=0.96\textwidth]{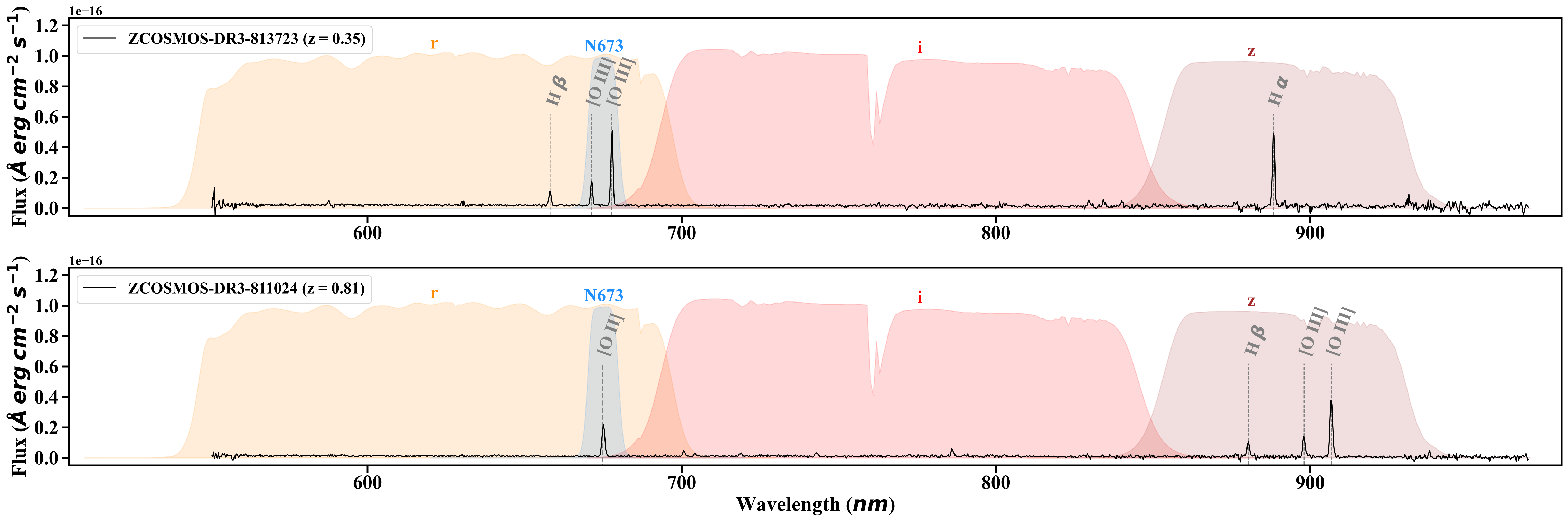}
\caption{Examples of ODIN-selected low-$z$ emission line galaxies superimposed on the filter transmission curves. The shaded curves represent the scaled filter transmission curves for the $r$, $i$, and $z$-band filters. In the top panel, the black line represents the spectrum for the $z = 0.35$ Green Pea-like galaxy ZCOSMOS-DR3-813723. The gray dashed vertical lines represent the expected locations of the H$\beta$, [\ion{O}{3}] doublet, and H$\alpha$ lines. Although the ZCOSMOS spectrum does not extend to lower wavelengths, this object's [\ion{O}{2}] emission line would fall within the $N501$ narrowband filter. In the bottom panel, the black line represents the spectrum for the $z = 0.81$ galaxy ZCOSMOS-DR3-811024. The gray dashed vertical lines represent the expected locations of the [\ion{O}{2}], H$\beta$, and [\ion{O}{3}] doublet lines.} 
\label{fig:combined_specs}
\end{center}
\end{figure*}

We also perform spectroscopic and photometric redshift cross-matches to our initial ($gi-N673$) selected LAE candidates. The spectroscopic cross-match confirms that the primary contaminants in our LAE candidate sample lie within a redshift range consistent with $z = 0.35$ [\ion{O}{3}] interlopers and in the region of our $grz$ color-color diagram where we predicted $z = 0.35$ [\ion{O}{3}] contamination. By visually inspecting the subset of these sources with accessible spectra \citep{lilly2007zcosmos, lilly2009zcosmos}, we find that they have similar emission line ratios to Green Pea-like $z = 0.35$ [\ion{O}{3}] emitters (see Figure \ref{fig:combined_specs}). Our photometric redshift cross match also shows high levels of contamination from sources with redshifts that are consistent with $z = 0.35$ [\ion{O}{3}] emitters in this same region in color-color space. Both cross-matches yield minimal contamination from sources with redshifts consistent with $z = 0.81$ [\ion{O}{2}] emitters. However, we place less weight on conclusions drawn from photometric redshifts due to their susceptibility to miss-classification of high EW emission line galaxies. These results suggest that $z = 0.81$ [\ion{O}{2}] emitters and $z = 0.35$ [\ion{O}{3}] emitters with high equivalent widths can be located in $grz$ color-color space, eliminated from ODIN's $z = 4.5$ LAE sample, and set aside for independent analysis. 


To further test our claims that the primary interloper contaminants in our sample of $z = 4.5$ LAE candidates are Green Pea-like $z = 0.35$ [\ion{O}{3}] emitters and that these interlopers preferentially reside in a specific region in $grz$ color-color space, we plot confirmed Sloan Digital Sky Survey (SDSS) Green Peas in the appropriate redshift range \citep{galaxy_zoo}. These Green Peas all have redshifts between 0.34 and 0.35 and correspond to objects with SDSS IDs 587732134315425958, 587739406242742472, and 587741600420003946 \citep{galaxy_zoo}. To place these Green Pea-like [\ion{O}{3}] emitters in $grz$ color-color space, we run their SDSS spectra through ODIN's filter set and obtain the flux density in each filter. This is accomplished using Equation \ref{eq:mock_obs}, where $f_{\nu}$ is the flux density, $S_{\lambda}$ is the galaxy's spectrum, $T$ is the filter transmission data, $c$ is the speed of light, and $\lambda$ is the wavelength.
\begin{equation}\label{eq:mock_obs}
    f_{\nu} = \frac{1}{c} \frac{\int S_{\lambda} T \lambda d \lambda}{\int T / \lambda d \lambda},
\end{equation}
We carry out these calculations by numerically integrating using Simpson's rule and then convert the flux density values $f_{\nu}$ into AB magnitudes. We find that all of these SDSS Green Peas reside in the predicted region of $grz$ color-color space (see Figure \ref{fig:specz_photoz_interlopers_cands}). 

As an additional check, we perform a similar analysis with a simulated $z = 0.35$ Green Pea-like galaxy spectrum and a simulated $z = 0.81$ [\ion{O}{2}] emitter spectrum. We create these simulated spectra using the stellar population synthesis package \texttt{FSPS} \citep[Flexible Stellar Population Synthesis; ][]{FSPS1, FSPS2}. For both simulations we use MESA Isochrones and Stellar Tracks (MIST) \citep[MIST;][]{MIST_2016_Dotter, MIST_2016_Choi, MIST_2011_Paxton, MIST_2013_Paxton, MIST_2015_Paxton}, the MILES spectral library \citep{MILES}, the DL07 dust emission library \citet{DL07}, a Salpeter IMF \citep{salpeter1955luminosity}, the Calzetti Dust law \citep{calzetti2000dust}, and turn on nebular emission and absorption in the intergalactic medium (IGM) absorption. Enabling nebular emission and IGM absorption tunes the stellar population to take on the properties of an observed galaxy. For the Green Pea-like galaxy, we also set the gas phase metallicity and the stellar metallicity parameters to $-1$. This metallicity adjustment fine-tunes the relative emission line strengths to match those of a typical Green Pea galaxy. For each spectrum, we compute the flux densities and AB magnitudes in ODIN's filter set using Equation \ref{eq:mock_obs}. We find that the simulated galaxies reside within both of their anticipated regions of $grz$ color-color space (see Figure \ref{fig:specz_photoz_interlopers_cands}). 

\subsubsection{[\ion{O}{2}] and [\ion{O}{3}] Emitter Selection Criteria}

\begin{figure}
\begin{center}
\includegraphics[width=0.48\textwidth]{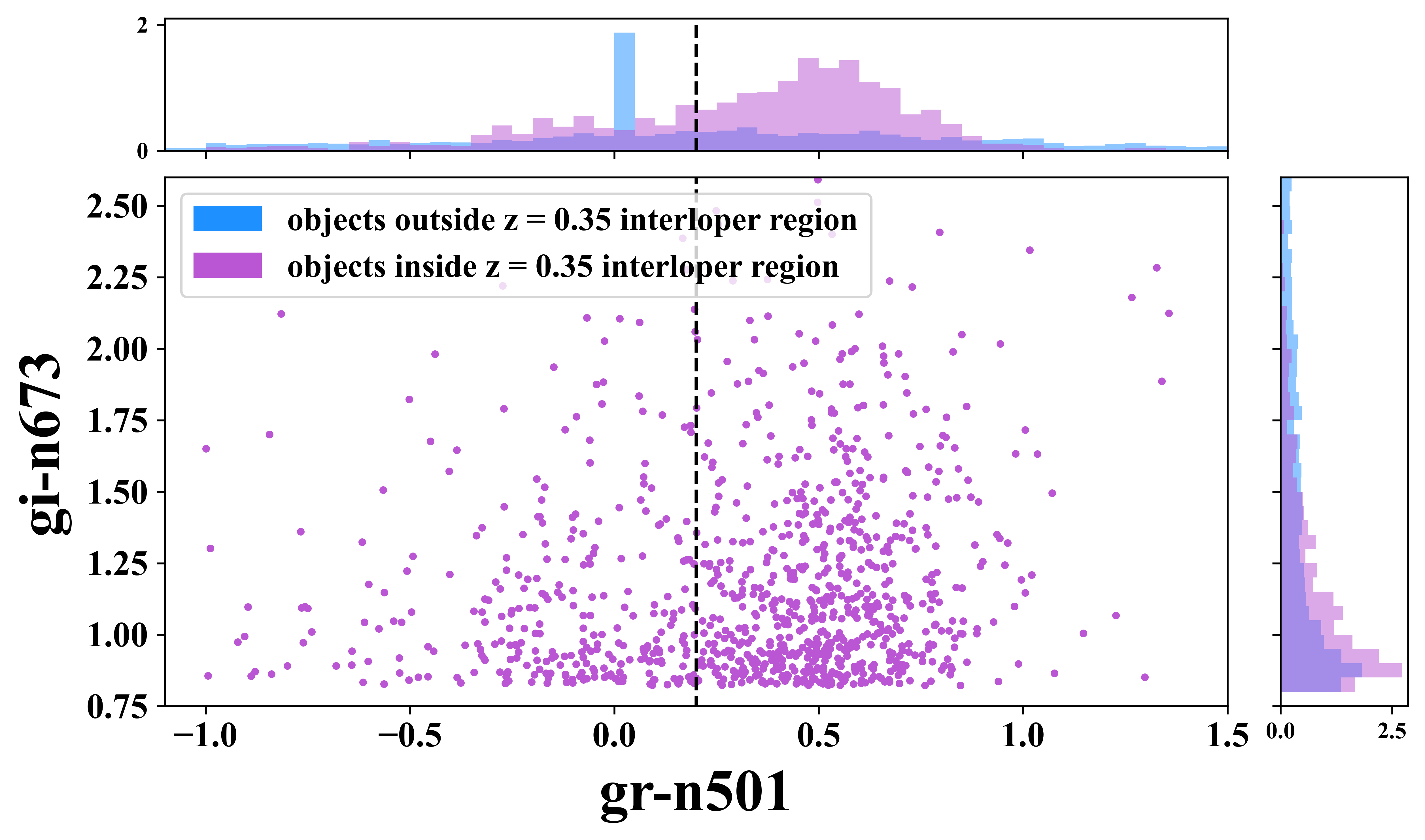}
\caption{Scatter-histogram illustrating $z = 0.35$ Green-Pea like galaxy selection diagnostics. The central plot is a ($gr-N501$)~vs.\ ($gi-N673$) color-color diagram of objects that passed our initial LAE selection criteria (see Section \ref{subsec:sel_criteria}) and reside inside the $grz$ $z = 0.35$ interloper region (see Figures \ref{fig:specz_photoz_interlopers} and \ref{fig:specz_photoz_interlopers_cands}), shown in purple. The upper histogram shows the ($gr-N501$) color distribution for candidates inside (purple) and outside (blue) the $grz$ $z = 0.35$ interloper region. (The peak in the blue histogram just above zero corresponds to relatively dim LAE candidates whose $N501$ flux is continuum-dominated.) The right plot shows the same for the ($gi-N673$) color. Each of the histograms is normalized such that its area sums to one. The dashed black vertical line shows the minimum $N501$ excess required for an object in the $grz$ $z = 0.35$ interloper region to be classified as a $z = 0.35$ Green-Pea like galaxy and be eliminated from the $z = 4.5$ LAE sample. Using the $grz$ $z = 0.35$ interloper region in tandem with this $N673$ and $N501$ narrowband excess diagnostic allows us to select a high-confidence sample of $z = 0.35$ Green-Pea like interlopers without removing the natural scatter in $N501$ narrowband excess in the LAE sample.} 
\label{fig:oiii_diagnostic}
\end{center}
\end{figure}

Our analyses show that the regions in $grz$ color-color space described in Subsection \ref{subsec:N673} are useful for targeting high EW $z = 0.35$ [\ion{O}{3}] emitter and $z = 0.81$ [\ion{O}{2}] emitter interlopers in our $N673$ dataset. We can see that the choice to carry out an ($gi-N673$) LAE candidate selection yields a remarkably low level of contamination from $z = 0.81$ [\ion{O}{2}] emitters. These analyses also reveal that the most prominent source of contamination in the initial ($gi-N673$) LAE candidate sample is $z = 0.35$ Green Pea-like [\ion{O}{3}] emitters. This discovery allows us to apply a specific and minimal LAE selection cut in $grz$ color-color space along with a ($gr - N501$) color excess criterion to eliminate bright $z = 0.35$ emission line galaxy interlopers (see Figure \ref{fig:oiii_diagnostic}). These additional cuts not only significantly enhance the purity of our $z = 4.5$ LAE sample, but also allow us to set aside this unique class of bright $z = 0.35$ Green Pea-like [\ion{O}{3}] emitters for future investigation. We therefore remove all sources that satisfy the additional criteria for our $z = 4.5$ LAE candidate selection and reserve the sources that do satisfy the following criteria for an [\ion{O}{3}] emitter candidate sample. 
\begin{enumerate} [leftmargin=1.3\parindent]

\item $0.4 \leq (g-r) \leq 0.85$
\item $0.05 \leq (r-z) \leq 0.55$
\item $(gr - N501) \geq 0.2 $

\end{enumerate}
At this point, we reject three additional spectroscopically confirmed low-redshift interlopers from our LAE sample with redshifts of 0.36, 0.37, and 0.80.

Supplementally, we can also generate a sample of $z = 0.81$ [\ion{O}{2}] emitter candidates by carrying out a ($r-N673$) selection and reserving objects that reside within the selected region in $grz$ color-color space defined by the below criteria. 
\begin{enumerate} [leftmargin=1.3\parindent]

\item $ -0.89 \leq (g-r) - 1.2(r-z) \leq -0.40$
\item $0.48 \leq (g-r) + 0.56(r-z) \leq 1.18$

\end{enumerate}

\subsection{Selection of $z = 3.1$ LAEs and $z = 0.35$ [\ion{O}{2}] Emitters}\label{subsec:N501}

For our $z = 3.1$ LAE selection, we carry out hybrid-weighted double-broadband continuum estimation using the $N501$, $g$-band, and $r$-band filters (see Figure \ref{fig:color_mag_comp} and Table \ref{tab:selection_results}). Following Table \ref{tab:weights}, we define the double-broadband continuum estimation $gr = 0.856 g + 0.144 r$. Since the [\ion{O}{3}] emission lines occur at rest-frame wavelengths of 495.9 nm and 500.7 nm, only very low-$z$ galaxies would have these emission lines at 501 nm. Because the EW distributions and luminosity functions of low redshift interlopers are lower at lower redshift, low redshift [\ion{O}{3}] emitters do not pose a threat to the purity of our $z = 3.1$ LAE sample. Additionally, due to their low redshifts, we expect most of these objects to be eliminated by the half-light radius cut. Therefore, the most likely source of low redshift interloper contamination is $z = 0.35$ [\ion{O}{2}] emitters. That being said, the EW of the [\ion{O}{2}] emission line tends to be significantly smaller than the corresponding [\ion{O}{3}] EW and the typical Ly$\alpha$ EW. 

In order to ensure that there is minimal contamination from $z = 0.35$ [\ion{O}{2}] emitters, we utilize the $N673$ narrowband filter, which is designed to pick up the [\ion{O}{3}] emission line for $z = 0.35$ galaxies (as discussed in the previous subsection). We find that the ($gi - N673$) color for the $z = 3.1$ LAE candidate sample is symmetrically distributed about a mean of $-0.013$. This shows that, as expected, if any [\ion{O}{2}] contaminants do exist, they are not also bright in [\ion{O}{3}]. We also find that the objects with higher ($gi - N673$) color for the original $z = 3.1$ LAE candidate sample do not preferentially reside in the region of $grz$ color-color space where we have previously identified the population of $z = 0.35$ [\ion{O}{3}] emitters in our $N673$-detected LAE sample.
These conclusions imply that our LAE candidate sample does not contain noticeable contamination from $z = 0.35$ [\ion{O}{2}] emitters, which is consistent with previous results from \citet{gronwall2007lyalpha}. Lastly, we remove three spectroscopically confirmed low-redshift interlopers from our LAE sample with redshifts 0.82, 2.10, and 2.14. We also confirm two LAE redshifts.

\subsection{Selection of $z = 2.4$ LAEs}\label{subsec:N419}

\begin{figure*}
\begin{center}
\includegraphics[width=1\textwidth]{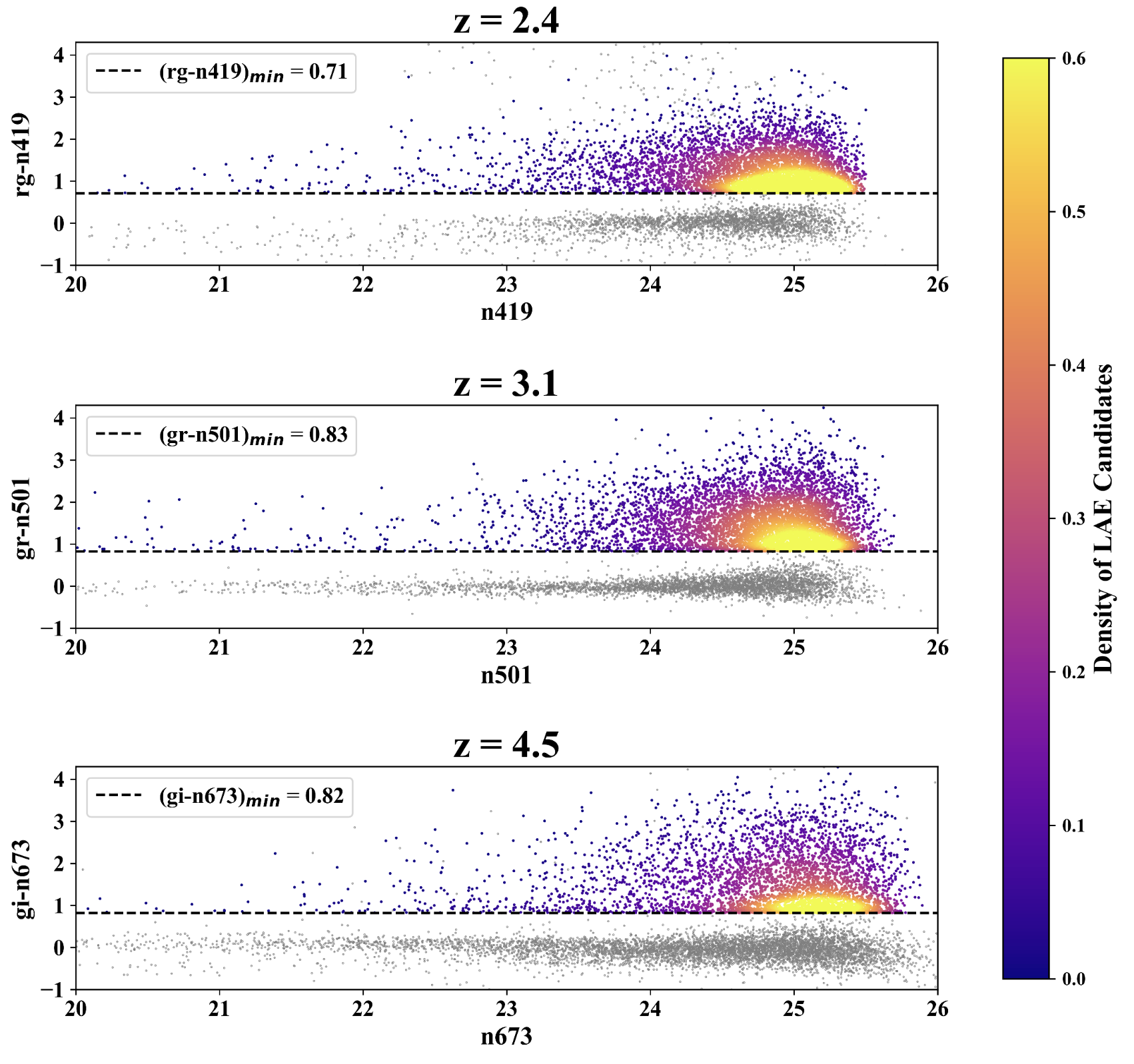}
\caption{Color-magnitude LAE selection diagrams for $z = 2.4$ (top), 3.1 (middle), and 4.5 (bottom). The x-axis represents the narrowband magnitude and the y-axis gives the narrowband excess as the difference between the double-broadband magnitude and the narrowband magnitude. The dashed line is the narrowband excess cut applied for each selection method. Random subsets of catalog sources are presented in gray. The color bar represents the density of LAEs in color-magnitude space.} 
\label{fig:color_mag_comp}
\end{center}
\end{figure*}

For our $z = 2.4$ LAE selection, we carry out hybrid-weighted double-broadband continuum estimation using the $N419$, the $g$-band, and $r$-band filters (see Figure \ref{fig:color_mag_comp} and Table \ref{tab:selection_results}). Following Table \ref{tab:weights}, we define the double-broadband continuum estimation $rg = -0.438 r + 1.438 g$.  Rather than using broadband filters on either side of the narrowband filter to estimate our continuum (i.e., $u$-band and $r$-band), we choose to use the $g$ and $r$ broadband filters to define the galactic continua. This is advantageous because it makes it possible to select $z = 2.4$ LAE candidates without direct use of the $u$-band filter, which is shallower than the $g$ and $r$-band filters. This choice is also beneficial because the $u$-band data cover a smaller area than the $g$ and $r$-bands, and are plagued by more systematic issues than the $g$ and $r$-bands.

Out of our three LAE candidate samples, the $z = 2.4$ LAE sample using our $N419$ filter is the least susceptible to low redshift emission line galaxy interlopers (with the exception of inevitable narrow and broad emission line AGNs). It is nonetheless important to complete a thorough spectroscopic follow-up on this candidate sample to fully assess its purity. Lastly, we reject 14 spectroscopically confirmed low-redshift interlopers from our LAE sample with redshifts 0.13, 0.13, 0.22, 0.79, 0.83, 0.88, 0.93, 1.02, 1.03, 1.16, 1.2, 1.35, 1.68, and 1.72. We also confirm seven LAE redshifts.







\section{Results}\label{sec:results}

\begin{figure*}
\begin{center}
\includegraphics[width=1\textwidth]{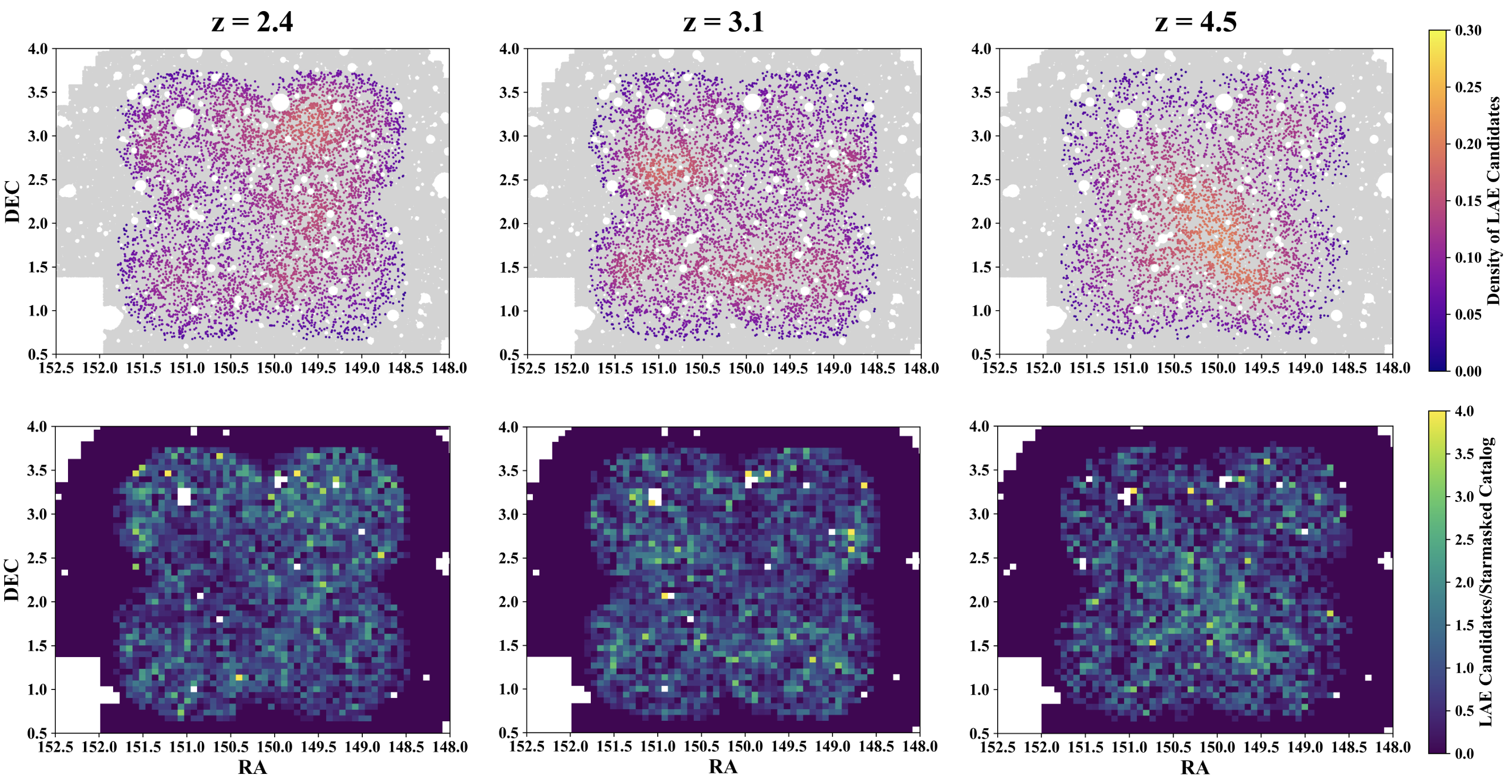}
\caption{Spatial distributions of $z = 2.4$ (left), 3.1 (middle), and 4.5 (right) LAE samples. The x-axis represents the right ascension (RA) in degrees and the y-axis represents the declination (DEC) in degrees for all panels. In the top three panels, the catalog sources are presented in gray and the LAE densities are represented by a Gaussian kernel density estimator and the colorbar. The bottom three panels are 2D histograms that represent the relative spatial distribution of LAEs. The colorbar demonstrates the ratio of LAE density to that of sources in the starmasked catalog, scaled to the average ratio in each dataset.} 
\label{fig:spatial_comp}
\end{center}
\end{figure*}

Using our selection criteria, we find samples of 6,032 $z = 2.4$ LAEs, 5,691 $z = 3.1$ LAEs, and 4,066 $z = 4.5$ LAEs in the extended COSMOS field ($\sim$9 deg$^2$; $\sim$7.8 deg$^2$ post-starmasking). The number of candidates remaining after each step in the LAE selection pipeline is presented in Table \ref{tab:selection_results}. The samples correspond to LAE densities of 0.21, 0.20, and 0.14 arcmin$^{-2}$, respectively. We also find that there are 665 $z = 0.35$ [\ion{O}{3}] emitters and 375 $z = 0.81$ [\ion{O}{2}] emitters. There are 8 spec$z$ matches in the $z = 0.35$ [\ion{O}{3}] emitter catalog and there is 1 spec$z$ match in the $z = 0.81$ [\ion{O}{2}] emitter catalog. All of these spec$z$ matches are in the corresponding redshift ranges. We present the color-magnitude LAE selection diagrams for all three redshifts in Figure \ref{fig:color_mag_comp}, where the LAEs are displayed in color and a sub-sample of random field objects are shown in gray. We present the spatial distribution of LAEs in each sample in Figure \ref{fig:spatial_comp}. The latter plots show that there are no pronounced systematic effects impacting our LAE selection as a function of spatial position at any of the three redshifts. The overdense regions in these figures also suggest that there are unique structures in the LAE candidate populations, providing the starting point for a subsequent clustering analysis (D. Herrera et al., in preparation).


\begin{table}[h!]
\centering
\caption{LAE selection summary statistics.}\label{tab:selection_results}
\begin{tabular}{lccc}
\hline
LAE Redshift & 2.4 & 3.1 & 4.5 \\
\hline\hline
Source extraction & 1,083,476 & 1,482,315 & 2,535,478 \\ 
Starmasking & 868,184 & 1,199,385 & 2,077,563 \\ 
Data quality cuts & 558,908 & 747,466 & 1,209,843 \\ 
LAE selection cuts & 6,100 & 5,782 & 4,870 \\ 
Interloper rejection & 6,046 & 5,694 & 4,069 \\
Spec$z$ rejection & 6,032 & 5,691 & 4,066 \\
\hline
Final LAE sample & 6,032 & 5,691 & 4,066 \\ 
\hline
\end{tabular}
\end{table}

\newpage

\subsection{Scaled Median Stacked SEDs}

Spectral Energy Distribution (SED) stacking is a technique used to represent generalized characteristics for a sample of objects. When creating a stacked SED, it is assumed that all galaxies in the sample have similar physical properties and that the properties of the stacked SED will match the physical properties of typical individual galaxies. As a consequence of this, every stacking method has the limitation that it cannot capture the diverse properties in a galaxy sample. However, SED stacking can be a helpful tool for understanding sample purity, especially for objects with faint continuum emission and expected continuum breaks such as LAEs. 

There are two primary classes of stacking: image stacking and flux stacking. Within each of these classes, there are three predominant stacking methods: mean, median, and scaled median. Mean stacking yields a good representative value if there are no outliers in the sample, but the result can be skewed if there is a wide spread in galaxy characteristics or contamination from AGNs or low$-z$ interlopers. Median stacking has less susceptibility to outliers and contaminants, but does not take into account the spectral shapes of all objects in a sample and is relatively inefficient. \citet{vargas2014stack} showed that the best simple stacking method for representing SED properties of $z$ = 2.1 LAEs is scaled median stacking, which has the added advantage that the influence of overall brightness variations is removed. In this study we choose to follow in the footsteps of \citet{vargas2014stack} and use flux scaled median stacking for our population SEDs. We outline the procedure for this method below.

In order to create scaled median stacked SEDs, we first find the median of the flux densities in our scaling filter $\tilde{f}_{scale}$. Then, we create a scaling factor for each source $\delta_i$ by computing the ratio of the median flux density in our scaling filter $\tilde{f}_{scale}$ to the flux density measurement for each source in our scaling filter $f_{scale, i}$. 
\begin{equation}
    \delta_i = \tilde{f}_{scale} / f_{scale, i}
\end{equation}
Next, we calculate the scaled flux density of a filter $[F_{filt}]$ by multiplying the flux density measurements in that filter $f_{filt, i}$ by the scaling factor $\delta_i$. 
\begin{equation}
    [F_{filt, i}] = f_{filt, i} \times \delta_i
\end{equation}
Lastly, we use the median of the scaled flux density for all sources in the filter to determine the filter's scaled median stacked flux density $\tilde{F}_{filt}$. By following this prescription for all of our filters, we create a scaled median stacked SED for each LAE sample. 

\begin{figure*}
\begin{center}
\includegraphics[width=\textwidth]{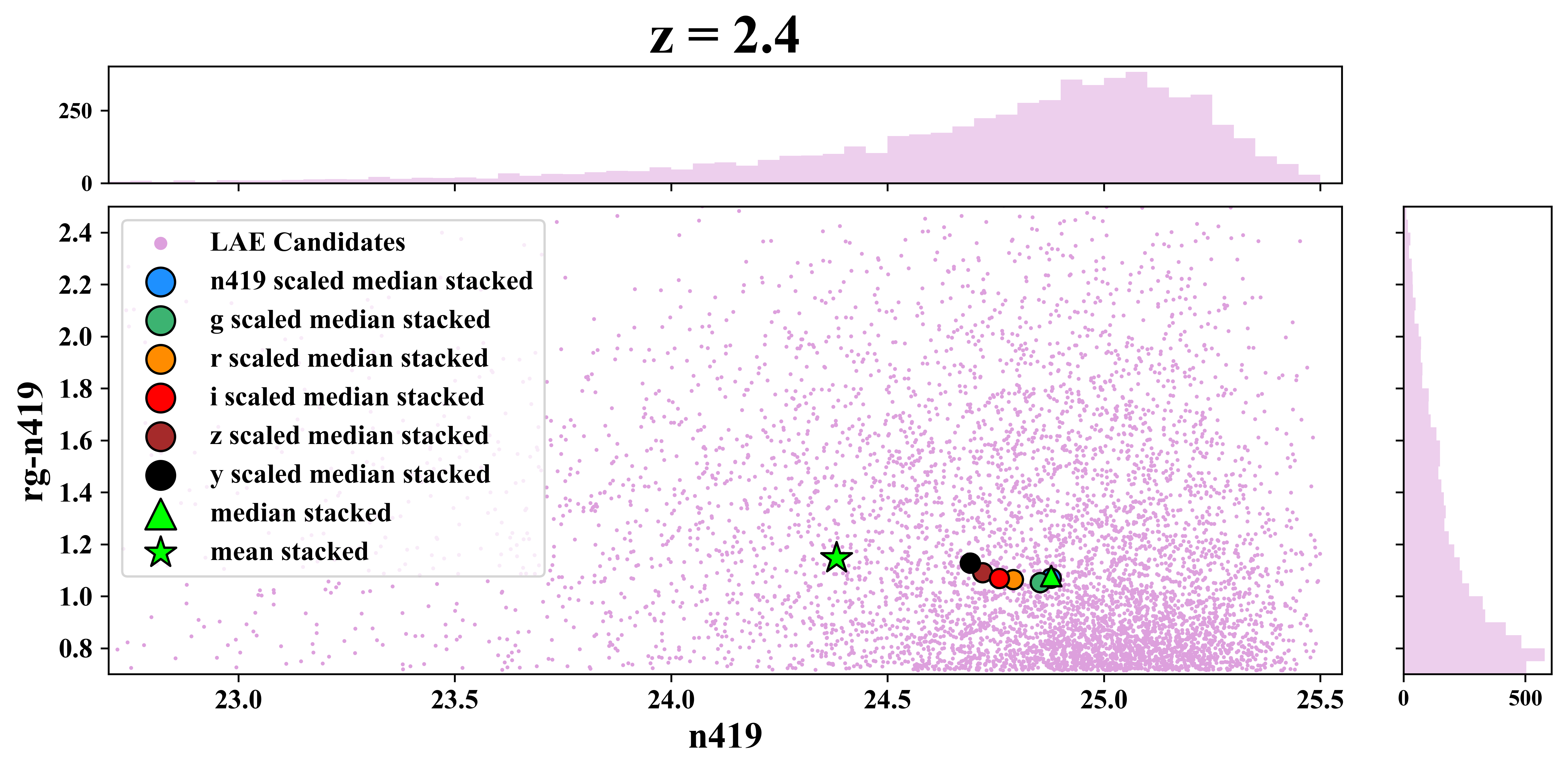}
\caption{Color-magnitude scatter-histogram for comparison of SED stacking methods using the $z = 2.4$ LAE sample. The x-axis represents the $N419$ narrowband magnitude and the y-axis represents the narrowband excess. The light pink represents the $z = 2.4$ LAEs, the colored circles represent the scaled median stacked LAEs, the green triangle represents the median stacked LAE, and the green star represents the mean stacked LAE in color-color space. The order of the circular points on the plot matches the legend from right to left. This figure demonstrates the general robustness of the scaled median stacking method and reinforces the conclusion that, of the three methods, this is the optimal stacking method for this analysis.} 
\label{fig:Cosmos_rg-n419-SE_i_stacked_SED_color_color_comp}
\end{center}
\end{figure*}

In addition to carrying out scaled median stacking, we also create median stacked and mean stacked SEDs for comparison. We find that the mean stacked SED yields flux densities that are much larger than for our (scaled) median stacked SEDs. This confirms that mean stacking is highly susceptible to outliers and brightness variations in our LAE sample. In contrast, we find that our scaled median stacked SEDs and our median stacked SEDs do not yield drastically different results, though our scaled median stacked SEDs have smaller interquartile ranges. We also find that our scaled median stacked SEDs are robust to changes in the scaling filter for all filters except for the $u$-band. This is not surprising. At $z$ = 3.1 and 4.5, the $u$ filter's bandpass lies partially or entirely blueward of the Lyman break, and even at $z$ = 2.4, the flux recorded by the filter is strongly affected by the Ly$\alpha$ forest. Large stochastic differences in $u$-band flux are therefore expected \citep[e.g.,][]{madau1995radiative, venemans2005properties}. 

We present the results of these stacked LAEs in color-magnitude space in Figure \ref{fig:Cosmos_rg-n419-SE_i_stacked_SED_color_color_comp}; $u$-band data have been excluded for the aforementioned reasons. Although Figure \ref{fig:Cosmos_rg-n419-SE_i_stacked_SED_color_color_comp} only shows the results for the $z = 2.4$ LAEs, the behavior is similar across all three redshifts. Overall, we find that the standard deviation in narrowband magnitude for the narrowband, $g$, $r$, $i$, $z$ and $y$ scaled median stacked LAEs is $0.04 \pm 0.01$ mag and the standard deviation in scaled median ($ab - NB$) color is $0.02 \pm 0.02$. The agreement among these values argues that our scaled median stacking methods are robust. These results also reinforce the conclusion that scaled median stacking is a defensible method for this analysis.





To form our LAE SEDs, we began by normalizing each galaxy's flux density to its measurement in the $i$-band; this filter does not contain a Ly$\alpha$ emission line nor any other strong spectral line feature at $z$ = 0.35, 0.81, 2.4, 3.1, or 4.5, and its use minimizes the interquartile ranges for our flux density values. Additionally, we exclude objects with $i$-band magnitude $\geq$ 40 (the low-flux density flag) from our SEDs since their small scaling factor causes the scaled flux densities of the other filters to become artificially inflated. We also do not include objects with no $u$-band data from the $u$-band stacks since the $u$-band covers a smaller area than the HSC filters used in the selection process. 

\begin{figure*}
\begin{center}
\includegraphics[width=\textwidth]{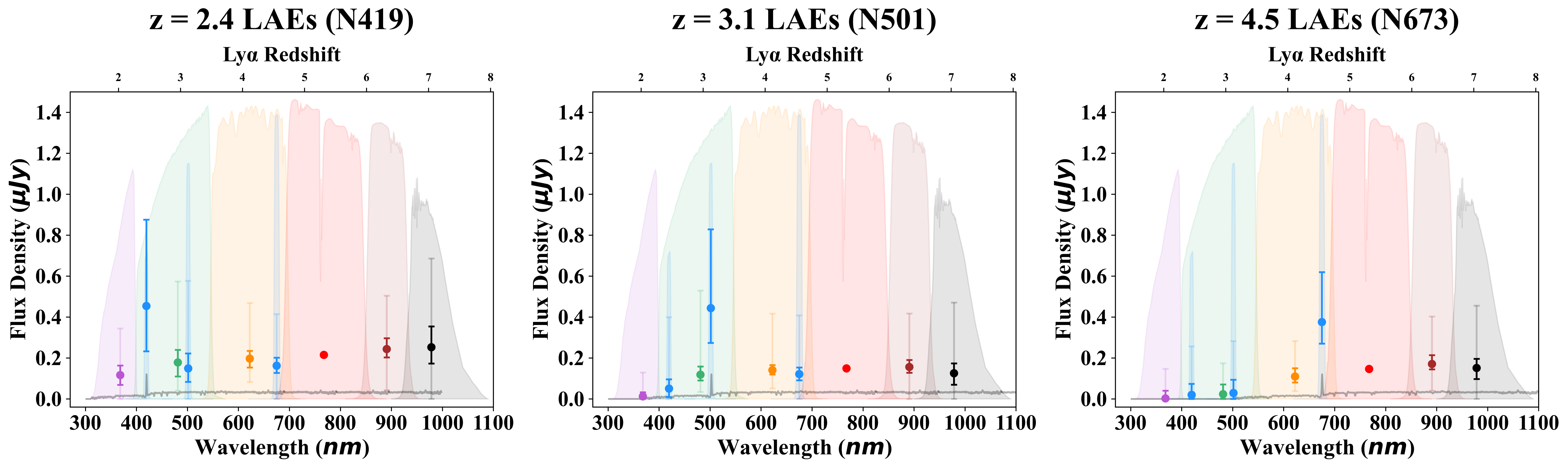}
\caption{$i$-band scaled median stacked spectral energy distributions for the LAE samples at $z = 2.4$ (left), 3.1 (middle), and 4.5 (right). The lower x-axis represents the wavelength in nanometers, the upper x-axis represents the the Ly$\alpha$ redshift, and the y-axis represents the flux density in microjansky. From left to right, the shaded curves represent the filter transmission for $u$, $g$, $r$, $i$, $z$, and $y$ broadband filters. The subplots also include filter transmission curves for $N419$, $N501$, and $N673$ narrowband filters, respectively. All the filter transmission curves have been scaled to fit the y-axis range of the flux density data. The $i$-band scaled median stacked flux density for each filter is plotted at the effective wavelength of the corresponding filter. The translucent error bars represent the 50\% confidence intervals and the opaque error bars represent the 95\% confidence intervals. (The $i$-band has no error bars since all data are scaled to this filter.) The subplots also include grayscaled LAE spectra as an aid to interpretation. The LAE spectra were adapted from \citet{shapley2003rest}. The strong presence of the Ly$\alpha$ forest decrement and the Lyman break in these SEDs suggests the success of our LAE candidate selection methods in achieving high sample purity.} 
\label{fig:scaled_median_stacked_sed}
\end{center}
\end{figure*}

We can assess the overall success of our LAE selection by examining the stacked SEDs. In Figure \ref{fig:scaled_median_stacked_sed}, we present the $i$-band scaled median stacked SEDs for the $z = 2.4$, 3.1, and 4.5 LAE candidate samples. These SEDs contain the key features that we expect to find in LAE spectra. Firstly, there is clear evidence for absorption by the Ly$\alpha$ forest in all three SEDs. The Ly$\alpha$ forest is characterized by absorption from hydrogen gas clouds in between the observer and the galaxy. This absorption occurs from the Ly$\alpha$ line down to shorter wavelengths, so we expect the Ly$\alpha$ forest decrement to occur most distinctly in the broadband whose effective wavelength is immediately below the effective wavelength of the corresponding narrowband. Our SEDs reveal that the Ly$\alpha$ forest decrement is present in the $u$-band for $z = 2.4$, in $N419$ at $z = 3.1$, and in the $r$-band at $z = 4.5$. We do not see a clear decrement in the $g$-band for the $z = 3.1$ SED because in this case the $g$-band also includes the Ly$\alpha$ emission line. We also find that the Lyman break is present in our SEDs. The Lyman break is characterized by the complete absorption of ionizing photons by gas below the short-wavelength end of the Lyman series transitions, the Lyman limit. In the rest-frame, this limit corresponds to 91.2 nm. At a redshift of 2.4, we expect the Lyman limit to occur at $\sim$310 nm. Because this wavelength falls out of the transmission ranges of our broadband filters, we do not see evidence for (or against) the Lyman limit in our $z = 2.4$ LAE candidate SED. At a redshift of 3.1, we expect the Lyman limit to occur at $\sim$374 nm. This is close to both the effective wavelength and long-wavelength limit of the $u$-band (see Table~\ref{tab:filters}). We see a strong effect from the Lyman break in the $u$-band for our $z = 3.1$ LAE SED. For the redshift 4.5 LAEs, we expect to find the Lyman limit at $\sim$502 nm. This is $\sim$20 nm longer than the effective wavelength and $\sim$50 nm shorter than the long-wavelength limit of the $g$-band (see Table \ref{tab:filters}). Therefore, in the $g$-band, $N419$, and $N501$ we see the partial effect of the Lyman break and in the $u$-band we see its full effect. Across the three redshifts, the strong presence of the Ly$\alpha$ forest decrement and the Lyman break suggests the general success of our LAE selections. 

\begin{figure*}
\includegraphics[width=\textwidth]{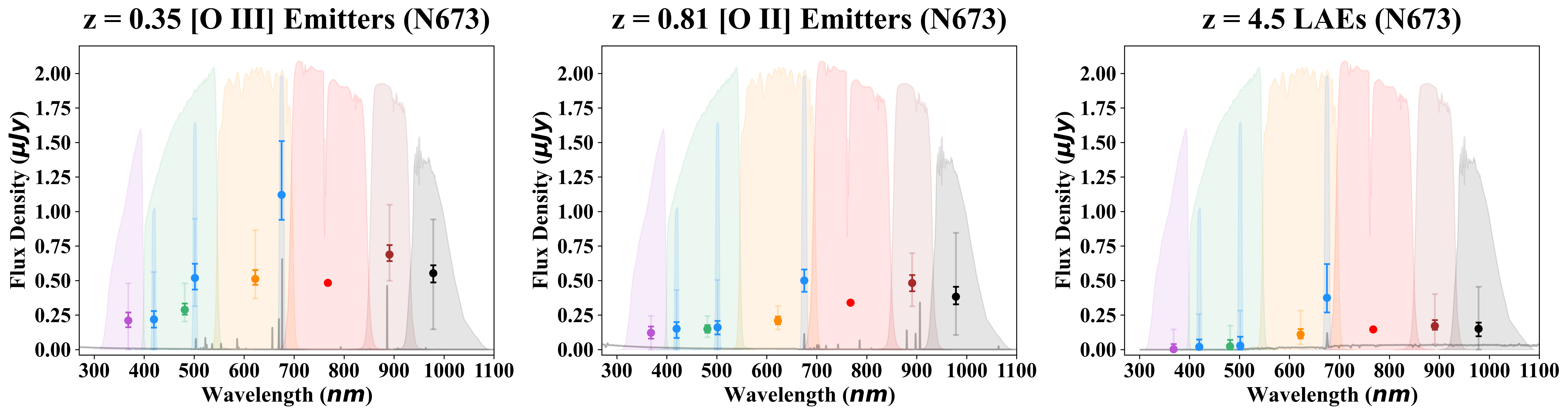}
\caption{$i$-band scaled median stacked spectral energy distributions for the N673-selected line emitters, namely the $z = 0.35$ [\ion{O}{3}] emitter sample (left), $z = 0.81$ [\ion{O}{2}] emitter sample (middle), and $z = 4.5$ LAE sample (right). The template of these plots follows Figure \ref{fig:scaled_median_stacked_sed}. [\ion{O}{3}] and [\ion{O}{2}] emitter spectra were generated using FSPS \citep{FSPS1, FSPS2}, and the LAE spectrum was adapted from \citet{shapley2003rest}. The lack of evidence for the Ly$\alpha$ forest decrement and the Lyman break in the [\ion{O}{3}] emitter and [\ion{O}{2}] emitter SEDs suggests that these populations are not dominated by $z = 4.5$ LAEs. Additionally, the elevated flux density in bands containing emission lines in the [\ion{O}{3}] emitter and [\ion{O}{2}] emitter spectra suggests that these populations are dominated by [\ion{O}{3}] emitters and [\ion{O}{2}] emitters, respectively.} 
\label{fig:N673_scaled_median_stacked_seds}
\end{figure*}

In Figure \ref{fig:N673_scaled_median_stacked_seds}, we also present the $i$-band scaled median stacked SEDs for the $z = 0.35$ [\ion{O}{3}] emitters, $z = 0.81$ [\ion{O}{2}] emitters, and $z = 4.5$ LAEs. Since all of these samples were selected from the $N673$-detected SE catalog, comparing them offers valuable insight into the success of our interloper rejection/selection methods. We find that the $z = 0.35$ [\ion{O}{3}] emitters and the $z = 0.81$ [\ion{O}{2}] emitters are generally much brighter in the $i$-band than $z=4.5$ LAEs; this is consistent with their much smaller luminosity distances. Additionally, the $z = 0.35$ [\ion{O}{3}] emitters have significant flux density in the $N501$ filter due to the presence of the redshifted [\ion{O}{2}] emission. Furthermore, we find that the Green-Pea like galaxies have heightened flux density in the $r$-band due to the presence of the H$\beta$, [\ion{O}{3}]$\lambda$4959, and [\ion{O}{3}]$\lambda$5007 emission lines and in the $z$-band due to the presence of the H$\alpha$ emission line (see Figure \ref{fig:combined_specs}). Similarly, the $z = 0.81$ [\ion{O}{2}] emitter systems have an excess of flux density in the $z$-band due to the presence of the H$\beta$, [\ion{O}{3}]$\lambda$4959, and [\ion{O}{3}]$\lambda$5007 emission lines (see Figure \ref{fig:combined_specs}). Lastly, we find that both the $z = 0.35$ [\ion{O}{3}] emitters and the $z = 0.81$ [\ion{O}{2}] emitters have significant emission in the $g$-band and $u$-band, whereas the $z = 4.5$ LAEs exhibit the presence of a partial and full Lyman break in these filters. These features imply that our rejection/selection methods for low-redshift emission line galaxy interlopers are successful. 

\subsection{Ly$\alpha$ Equivalent Width Distributions}\label{subsec:ew_estimations}

\begin{figure*}
\begin{center}
\includegraphics[width=\textwidth]{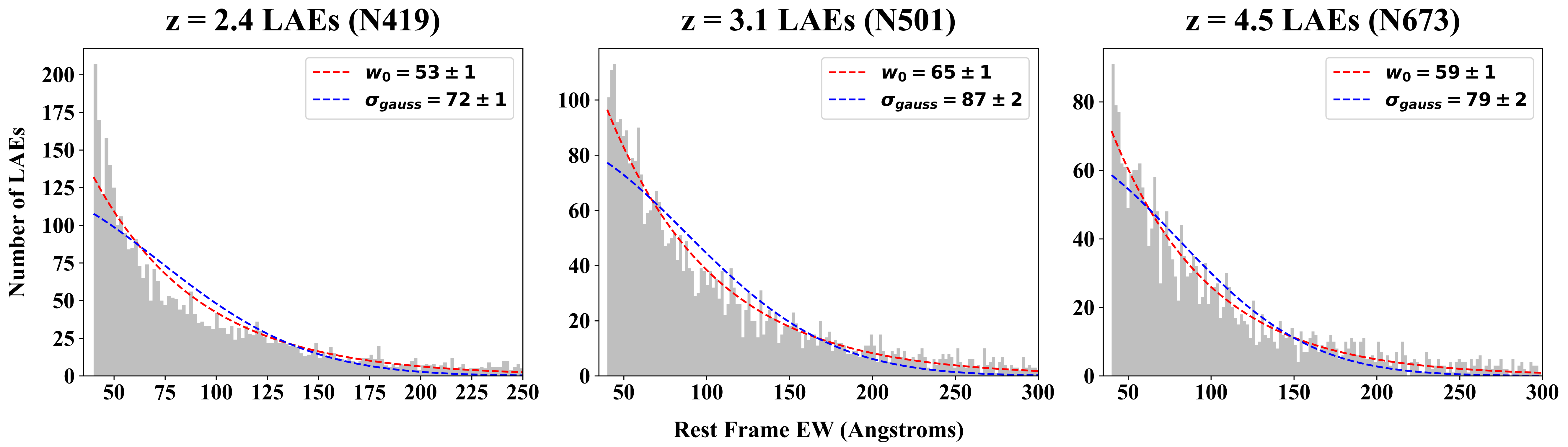}
\caption{Rest-frame Ly$\alpha$ equivalent width distributions for $z = 2.4$ (left), 3.1 (middle), and 4.5 (right) LAE samples. The x-axis represents the rest-frame Ly$\alpha$ EW in angstroms and the y-axis represents the number of LAEs in a given EW bin. A histogram of the EW distribution for each sample is presented in gray. The scale length for an exponential fit $w_0$ to each distribution is presented in angstroms with a red dashed line. The scale length for a Gaussian fit $\sigma_{gauss}$ to each distribution is presented in angstroms with a blue dashed line. These results suggest an evolution in scale length with increasing redshift.} 
\label{fig:Lya_ew_dists}
\end{center}
\end{figure*}

Now that we have shown our LAE samples have high levels of purity, we can use them to quantify the Ly$\alpha$ Equivalent Width (EW) distribution at each redshift. We define the EW as the width of a rectangle from zero intensity to the continuum level with the same area as the area of the emission line. Physically, the Ly$\alpha$ EW is related to the burstiness of LAEs since it compares the Ly$\alpha$ emission from O and B stars to the continuum emission from O, B, and A stars (with radiative transfer) \citep{broussard2019star}. Therefore, quantifying the Ly$\alpha$ EW distribution is helpful for comparing sample characteristics of LAEs.

We derive the rest-frame Ly$\alpha$ $EW$ distribution for each LAE sample following the methodologies of \citet{venemans2005properties} and \citet{guaita2010lyalpha}. For a detailed derivation, see the Appendix of \citet{guaita2010lyalpha}. 

First, we take the rest-frame equivalent width as $EW_{obs}/(1+z)$, where the observed equivalent width $EW_{obs}$ is defined as follows:
\begin{equation}
    EW_{obs} = A/B,
\end{equation}
where $A$ and $B$ are defined in Equations \ref{eq:A} and \ref{eq:B}. 
\begin{equation}\label{eq:A}
    A = Q_{NB} - Q_{ab} 10^ {(ab - NB)/2.5}
\end{equation}

$$B = \frac{w_{BB} T_{EL,BB} (c/\lambda_{EL}^2) 10^{(ab - NB)/2.5 }} {\int (c/\lambda^2) T_{BB}(\lambda) d\lambda}$$
\begin{equation}
    \label{eq:B}
    - \frac{T_{EL,NB} (c/\lambda_{EL}^2)} {\int (c/\lambda^2) T_{NB}(\lambda) d\lambda}
\end{equation}
In Equation \ref{eq:A}, $Q$ is the fraction of the continuum flux in a particular filter that is transmitted by the Ly$\alpha$ forest, $ab$ is the double-broadband magnitude and $NB$ is the narrowband magnitude. We define $Q$ using Equation \ref{eq:Q}, 
\begin{equation}\label{eq:Q}
    Q_{filt} = \frac{ \int e^{-\tau_{eff}(\lambda)} (c/\lambda^2) T(\lambda) d\lambda} {\int (c/\lambda^2) T(\lambda)  d\lambda},
\end{equation}
where $T(\lambda)$ is the filter transmission at a given wavelength and $\tau_{eff}(\lambda)$ is the effective opacity of HI. For this analysis, we use Equation \ref{ew:tau} as an approximation for all observed wavelengths below the redshifted Ly$\alpha$ line \citep{press1993properties, madau1995radiative, venemans2005properties}. 
\begin{equation}\label{ew:tau}
    \tau_{eff}(\lambda) = 0.0036 \left( \frac{\lambda}{1216 \text{ \AA}} \right) ^{3.46}
\end{equation}
In Equation \ref{eq:B}, $BB$ refers to the selection broadband that also has a flux contribution from the emission line and $w_{BB}$ is the weight assigned to that broadband. For the ($rg$-$N419$) $z = 2.4$ LAE selection and the ($gr$-$N419$) $z = 3.1$ LAE selection, this broadband corresponds to the $g$-band. In the case of the ($gi$-$N673$) $z = 4.5$ LAE selection, neither of the broadband filters has a flux contribution from the emission line, so the first term in Equation \ref{eq:B} vanishes entirely. $T_{EL}$ is obtained by averaging the filter transmission over the narrowband filter transmission curve, which is used as a proxy for the LAE redshift probability distribution function. This is justifiable since the filter transmission curve is close to a top hat. $\lambda_{EL}$ is the wavelength corresponding to the emission line, i.e., the narrowband effective wavelength. 

We fit the resulting Ly$\alpha$ EW distributions using an exponential distribution as shown in Equation \ref{eq:exp} and a Gaussian distribution as shown in Equation \ref{eq:gauss}, where $N$ is the number of LAEs in a given EW bin, $C$ is a constant of the fit, $EW$ is the rest-frame Ly$\alpha$ EW, and $w_0$ and $\sigma_{gauss}$ are the respective scale lengths in angstroms. 
\begin{equation}\label{eq:exp}
    N = C \exp{\left(\frac{-EW}{w_0}\right)}
\end{equation}
\begin{equation}\label{eq:gauss}
    N = C \exp{\left(\frac{-EW^2}{2\sigma_{gauss}^2}\right)}
\end{equation}

We present these Ly$\alpha$ EW distributions and fits for the $z = 2.4$, 3.1, and 4.5 LAE samples in Figure \ref{fig:Lya_ew_dists}. To obtain a robust fit, we choose to clip our distributions at a minimum EW of 40 {\AA}. 
We also choose to exclude objects with EW above 400 {\AA} since the highest equivalent widths are associated with galaxies that are extremely faint in the continuum, and thus poorly measured. This results in the exclusion of less than 1\% of our LAE sample. Lastly, we choose to use 200 bins for the fits, corresponding to the minimum bin number for which the scale lengths for all three datasets become stable. We find that the exponential scale lengths for the three LAE samples are $w_0$ = 53$\pm$1, 65$\pm$1, and 59$\pm$1 {\AA}; and the Gaussian scale lengths are $\sigma_{gauss}$ = 72$\pm$1, 87$\pm$2, and 79$\pm$2 {\AA}, respectively. The reduced $\chi^2$ values for the exponential scale lengths are 1.9, 1.1, and 1.3; and the reduced $\chi^2$ values for the Gaussian scale lengths are 3.9, 2.6, and 2.3, respectively. The reduced $\chi^2$ values unanimously show preference toward an exponential fit when compared to a Gaussian fit. Although there is significant variation in the literature results, the $w_0$ scale lengths are similar to previous findings and the $\sigma_{gauss}$ are between $\sim10-40$\% lower than most previous findings \citep[e.g.,][]{gronwall2007lyalpha, ouchi2008subaru, nilsson2009evolution, guaita2010lyalpha, ciardullo2011evolution, kerutt2022equivalent}. Contamination from low-redshift emission-line or continuum-only galaxies tends to reduce the scale length; contamination from spurious objects tends to increase them, since the EWs are formally infinite when an ``object'' is only luminous in the narrow-band. We have made a careful effort to avoid all of these types of contamination, with our stacked SED analysis providing evidence against significant low-redshift contamination. Our finding that the EW scale length is at the lower end of results in the literature might indicate that we are free from significant contamination from spurious objects. However, it is possible that these results are impacted by selection effects. For example, some studies have shown that excluding objects with fainter UV continuum can result in smaller scale lengths \citep[e.g.,][]{oyarzun2017comprehensive, hashimoto2017muse}. Obtaining a better understanding of contamination rates and sample characteristics from each type of interloper will further improve the accuracy of these measurements. Additionally, the full ODIN LAE sample should result in even more precise measurements of the Ly$\alpha$ $EW$ distributions.

\subsection{LAEs with Measured $EW \geq 240$ {\AA}}\label{subsec:high_ew_laes}

Additionally, we investigate the objects with $EW \geq 240$ \AA. It has been speculated that a real LAE with $EW$ in this regime could have a normal stellar population with a clumpy dust distribution or could be composed of young, massive, metal-poor stars or Population III stars; however, measurements of the short-lived \ion{He}{2} $\lambda$1640 line and \ion{C}{4} $\lambda$1549 render the true composition of these systems ambiguous \citep{kashikawa2012lyalpha}. We find that there are 484, 561, and 245 LAEs in this regime at $z$ = 2.4, 3.1, and 4.5, respectively. 

We seek to understand the likelihood that these objects are real and are not the result of noise. In order to accomplish this, we first truncate our $EW$ distributions at 240 $\text{\AA}$ and forward-model the scatter in our data using a bootstrapping method to see how many objects exceed a $EW$ of 240 $\text{\AA}$. We accomplish this by taking each observed object with $EW < 240$ $\text{\AA}$ and applying random values within 1$\sigma$ of that object's noise in the double-broadband magnitude and in the narrowband magnitude, then recalculating $EW$. We then carry out this process multiple times until the average fraction of objects above 240 $\text{\AA}$ converges. Using this method, we find that $10\%^{+0.4}_{-0.5}$, $31\%^{+2}_{-0.3}$, and $191\%^{+1}_{-5}$ of the objects above 240 $\text{\AA}$ can be explained by noise, respectively. However, we find that objects with $EW \geq 240$ $\text{\AA}$ tend to have higher noise in their double-broadband magnitudes than objects with $EW < 240$ $\text{\AA}$. In order to account for this, we apply a similar noisification method where we instead take each observed object with $EW < 240$ $\text{\AA}$ and apply random noise values from the high $EW$ sample to the double-broadband magnitude and the narrowband magnitude, then recalculate $EW$. Using this method, we find that $102\%^{+0.2}_{-4}$, $85\%^{+4}_{-0.05}$, and $411\%^{+1}_{-8}$ of the objects above 240 $\text{\AA}$ can be explained by noise, respectively. Although the bootstrapping method suggests that there may be objects with truly high $EW$ in the z = 2.4 and 3.1 samples, the latter method implies that the high $EW$ objects might be explained by the large fraction of the sample that is formally undetected in the broad-band imaging, leading to large uncertainties in $EW$. Additionally, we carry out Gaussian error propagation and find that only 1.0\%, 1.2\%, and 1.2\% of these high $EW$ objects have measured rest-frame $EW>240${\AA} at 3$\sigma$ significance, respectively. Follow-up spectroscopy of this subset will be a priority to better understand how many of these objects truly have $EW \geq 240$ $\text{\AA}$.

\section{Conclusions and Future Work}\label{sec:conc}

ODIN is a NOIRLab survey program designed to discover LAEs by combining data taken through three narrowband filters custom-made for the Blanco 4-m telescope's DECam imager \citep{odin_survey} with archival broadband data from the HSC and CLAUDS. ODIN's narrowband filters, $N419$, $N501$, and $N673$, allow us to identify samples of LAEs at redshifts 2.4, 3.1, and 4.5, corresponding to epochs 2.8, 2.1, and 1.4 Gyrs after the Big Bang, respectively. When the ODIN survey is complete, we expect to discover $>$100,000 LAEs in seven of the deepest wide-imaging fields up to a narrowband magnitude of $\sim$25.7 AB, covering an area of $\sim$100 deg$^2$. 

In this paper, we used data from ODIN's first completed field covering $\sim$9 deg$^2$ in COSMOS to introduce innovative techniques for selecting LAEs and other samples of emission line galaxies using narrowband imaging. These include LAE samples at $z = 2.4$, 3.1, and 4.5, as well as samples of $z = 0.35$ [\ion{O}{3}] emitters and $z = 0.81$ [\ion{O}{2}] emitters. The main conclusions of this work are summarized below. 

\begin{enumerate} [leftmargin=1.3\parindent] 

    \item We developed a narrowband LAE selection method that utilizes a new technique to estimate emission line strength, the \textit{hybrid-weighted double-broadband continuum estimation} technique. Using this technique, we treated sources with S/N $\geq$ 3 in both single broadbands by assuming a power law SED and treated sources with S/N $<$ 3 in either broadband by assuming a linear spectral slope. This technique allowed us to better estimate expected continuum emission at the location of each narrowband filter by utilizing data from \textit{any} two nearby broadbands. This method provided the flexibility to choose optimal broadband filters that maximize the data area and quality and to avoid broadbands that may be heavily impacted by features in low redshift emission line interlopers. 

    \item Utilizing this new technique, we performed $z = 2.4$, 3.1, and 4.5 LAE candidate selections in the extended COSMOS field using broadband data from the HSC and narrowband data collected with DECam. We used the $N419$, $r$, and $g$-bands for our initial $z = 2.4$ LAE selection; the $N501$, $g$, and $r$ bands for our initial $z = 3.1$ LAE selection; and the $N673$, $g$, and $i$ bands for our initial $z = 4.5$ LAE selection. 

    \item We found that the main source of low redshift emission line contamination in our LAE samples was very bright $z = 0.35$ Green Pea-like galaxies. Our data also revealed that these galaxies occupy a compact and distinct region of $grz$ color-color space. Moreover, since the ODIN survey was designed in anticipation of $z = 0.35$ contaminants, the filter bandpasses were designed to ensure that the majority of $z = 0.35$ emission line galaxies will have [\ion{O}{3}] emission in the $N673$ narrowband filter and [\ion{O}{2}] emission in the $N501$ narrowband filter. Despite having emission lines detectable in both the $N673$ and the $N501$ narrowband filters, our results suggested that these $z = 0.35$ bright Green Pea-like galaxies are only a strong source of contamination in our $N673$ $z = 4.5$ LAE selection. By taking advantage of the $grz$ color criteria and the estimated $N673$ and $N501$ excess flux densities, we were able to identify and set aside a sample of 665 $z = 0.35$ Green Pea-like objects for further analysis. Although we did not find that $z = 0.81$ [\ion{O}{2}] emitters are a notable source of contamination in our $z = 4.5$ LAE candidate sample, we found that they also occupy a compact and distinct region of $grz$ color-color space and are selectable using the $N673$ and $r$-band filters. Thus, we also set aside a sample of $z = 0.81$ [\ion{O}{2}] emitter galaxies for future analysis. 
    
    \item We found that there are 6,032, 5,691, and 4,066 LAEs at $z = 2.4$, 3.1, and 4.5, respectively, in the extended COSMOS field ($\sim$9 deg$^2$). The samples imply LAE surface densities of 0.21, 0.20, and 0.14 arcmin$^{-2}$, respectively. These results were in agreement with the predictions outlined in \citet{odin_survey}. We also defined samples of 665 $z = 0.35$ Green Pea-like galaxies and 375 $z = 0.81$ [\ion{O}{2}] emitters. 

    \item We developed $i$-band flux density scaled median stacked SEDs for the $z = 2.4$, 3.1, and 4.5 LAE samples as well as the $z = 0.35$ Green Pea-like [\ion{O}{3}] emitter and $z = 0.81$ [\ion{O}{2}] emitter galaxy contaminants. We found that our $z = 2.4$, 3.1, and 4.5 LAE SEDs display clear features that are unique to LAEs such as the Ly$\alpha$ forest decrement and Lyman break.  We found that our $z = 0.35$ Green Pea-like [\ion{O}{3}] emitter and $z = 0.81$ [\ion{O}{2}] emitter SEDs have features unique to their respective populations. Our stacked SEDs revealed broad consistency in each sample, implying that our samples have high levels of purity. 

    \item We calculated Ly$\alpha$ equivalent width distributions for the $z = 2.4$, 3.1, and 4.5 LAE samples. We found that the EW distributions are best fit by exponential functions with scale lengths of $w_0$ = 53$\pm$1, 65$\pm$1, and 59$\pm$1 {\AA}, respectively. These scale lengths are at the lower end of the values reported in the literature. The precision of these measurements should improve for the considerably larger LAE sample expected from the full ODIN survey. 

    \item We found that an impressive $\sim 10$\% of our LAE samples have measured rest-frame equivalent width $\geq 240$ $\text{\AA}$, providing possible evidence of nonstandard IMFs or clumpy dust. However, deep spectroscopic follow-up is needed to ascertain how many of these equivalent widths are real as opposed to noise due to low continuum S/N.

\end{enumerate}

ODIN's LAE samples will allow us to quantify the temporal evolution of LAE clustering properties, bias, dark matter halo masses, and halo occupation fractions (D. Herrera et al., in preparation). As HETDEX and DESI-II work to probe dark energy using LAEs, ODIN's improved understanding of which dark matter halos host LAEs can allow these groups to better simulate their systematics, and will have a direct impact on their measurements of cosmological constraints. Furthermore, ODIN's LAE sample will allow us to uncover properties of individual LAEs such as their stellar mass, star formation rate, dust attenuation, timing of stellar mass assembly, and the processes of star formation and quenching. Once completed, this work will help us to better understand the relationship between LAEs, their present-day analogs, and their primordial building blocks. 

\section{Acknowledgements}


This work utilizes observations at Cerro Tololo Inter-American Observatory, NSF’s NOIRLab (Prop. ID 2020B-0201; PI: K.-S. Lee), which is managed by the Association of Universities for Research in Astronomy under a cooperative agreement with the National Science Foundation.

This material is based upon work supported by the National Science Foundation Graduate Research Fellowship Program under Grant No. DGE-2233066 to NF. NF and EG would also like to acknowledge support from NASA Astrophysics Data Analysis Program grant 80NSSC22K0487 and NSF grant AST-2206222. NF would like to thank the LSSTDA Data Science Fellowship Program, which is funded by LSST Discovery Alliance, NSF Cybertraining Grant 1829740, the Brinson Foundation, and the Moore Foundation; her participation in the program has benefited this work greatly.  
KSL and VR acknowledge financial support from the
National Science Foundation under Grant No. AST-2206705 and from the Ross-Lynn Purdue Research Foundation Grant. 
BM and YY are supported by the Basic Science Research Program through the National Research Foundation of Korea funded by the Ministry of Science, ICT \& Future Planning (2019R1A2C4069803).
LG 
and AS thank support from FONDECYT regular proyecto No. 1230591. 
HS acknowledges the support of the National Research Foundation of Korea grant, No. 2022R1A4A3031306, funded by the Korean government (MSIT). 
The Institute for Gravitation and the Cosmos is supported by the Eberly College of Science and the Office of the Senior Vice President for Research at the Pennsylvania State University. 

We thank Masami Ouchi for helpful comments on this paper. We also thank the anonymous referee for their thoughtful suggestions to improve this work.


\bibliography{sample631}{}
\bibliographystyle{aasjournal}

\end{document}